\documentclass[fleqn,usenatbib]{mnras}

\usepackage{newtxtext,newtxmath,subcaption,float}
\usepackage[T1]{fontenc}

\DeclareRobustCommand{\VAN}[3]{#2}
\let\VANthebibliography\thebibliography
\def\thebibliography{\DeclareRobustCommand{\VAN}[3]{##3}\VANthebibliography}

\newcommand{\Xlan}{X$_{\rm lan}$}
\newcommand{\vmax}{$v_{\rm max}$}

\newif\ifcomments
\commentstrue % set to \commentsfalse to hide comments

\usepackage{graphicx}	% Including figure files
\usepackage{amsmath}	% Advanced maths commands

\title{Free Neutron Decay in Kilonova Ejecta: X-ray/UV Flashes with Non-Thermal Effects}

\author[D. Brethauer et al.]{
Daniel Brethauer,$^{1,2}$\thanks{E-mail: daniel\_brethauer@berkeley.edu}
Daniel Kasen,$^{1,2,3,4}$
Smaranika Banerjee$^{5}$,
\newauthor
Raffaella Margutti$^{1,2,3}$,
and Ryan Chornock$^{1,2}$
\\
$^{1}$Department of Astronomy, University of California, Berkeley, CA 94720-3411, USA\\
$^{2}$Berkeley Center for Multi-messenger Research on Astrophysical Transients and Outreach (Multi-RAPTOR), University of California, Berkeley, CA 94720-3411, USA\\
$^{3}$Department of Physics, University of California, 366 Physics North MC 7300, Berkeley, CA 94720, USA\\
$^{4}$Nuclear Science Division, Lawrence Berkeley National Laboratory, 1 Cyclotron Rd, Berkeley, CA, 94720, USA\\
$^{5}$ {Astrophysics sub-Department, Department of Physics, University of Oxford, Keble Road, Oxford OX1 3RH, UK}
}

\date{Accepted XXX. Received YYY; in original form ZZZ}

\pubyear{\the\year{}}

\usepackage{amsfonts,amsmath,natbib,color,gensymb,longtable,verbatim,graphicx,float,soul,multirow}

\begin{document}
\label{firstpage}
\pagerange{\pageref{firstpage}--\pageref{lastpage}}
\maketitle

% Abstract of the paper
\begin{abstract}
Neutron star-bearing compact-object mergers can create a kilonova, a transient powered by the radioactive decay of material synthesized via rapid neutron capture ($r$-process). During the merger, a fraction of the ejecta is launched at high speeds ($\gtrsim$ 0.4c) that can lead to neutrons evading capture onto seed nuclei, resulting in free neutrons. The free neutrons then decay and inject additional energy that power early-time ($\lesssim$1 day) emission. Here, using \texttt{Sedona}, we present a grid of the first non-local thermodynamic equilibrium and frequency-dependent opacity radiative transfer simulations to predict early X-ray/UV/optical emission from free neutron decay that span the expected theoretical range of free neutron mass, mixing with $r$-process material, and extent of the high-velocity ejecta tail. The emission can be characterized by an SED that rapidly shifts from X-rays of $\sim\rm{few} \times10^{41}-10^{42}$ erg s$^{-1}$ in the first $\sim$ minutes to a far-UV and near-UV peak on the timescale of $\sim$ minutes to hours, followed by enhanced optical and IR emission for sufficiently large free neutron masses. The properties of the free neutron ejecta are most distinguishable, in principle, at extreme-UV and far-UV wavelengths, with SED peak flux and wavelength determined by the maximum velocity of the ejecta and the mixing. In the far-UV and near-UV, free neutron ejecta masses as small as 10$^{-7}$ M$_\odot$ exhibit a unique bump compared to neutron-free models, demonstrating the importance of upcoming UV missions like UVEX and ULTRASAT to constraining the nucleosynthetic environment of neutron star mergers.
%Additionally, we simulate one 2D model and find that there is time-dependent polarization (\textcolor{green}{number to number}) and viewing angle dependence \textcolor{green}{that shows brighter emission at the poles??}.

\end{abstract}

% Select between one and six entries from the list of approved keywords.
% Don't make up new ones.

%%%%%%%%%%%%%%%%%%%%%%%%%%%%%%%%%%%%%%%%%%%%%%%%%%

%%%%%%%%%%%%%%%%% BODY OF PAPER %%%%%%%%%%%%%%%%%%

\section{Introduction} \label{sec:intro}

Compact-object mergers involving a neutron star (NS) have long been at the forefront of high-energy phenomena, from short gamma ray bursts (GRBs; e.g., \citealt{Eichler89,Narayan92}) to sites of \textit{rapid neutron capture} ($r$-process) nucleosynthesis \citep{Lattimer&Schramm74,Lattimer&Schramm76,Symbalisty&Schramm82,Eichler89,Freiburghaus99,Rosswog99}. The discovery of the gravitational wave source GW\,170817 and its electromagnetic counterparts, AT\,2017gfo and GRB\,170817A \citep{Abbott17,Andreoni17,Arcavi17,Chornock17,Coulter17,Cowperthwaite17,Diaz17,Drout17,Evans17,Goldstein17,Hu17,Kasliwal17,Lipunov17,Pian17,Savchenko17,Shappee17,Smartt17,Soares-Santos17,Tanvir17,Utsumi17,Valenti17,Pozanenko18}, confirmed the connection between compact-object mergers and some short GRBs, ushering in a new era transient multi-messenger astronomy with gravitational waves (see \citealt{Nakar20} and \citealt{Margutti&Chornock21} for reviews). The kilonova (KN) associated with GW\,170817, AT\,2017gfo, was a multi-wavelength transient powered by the radioactive decay of $r$-process material. The detected emission of AT\,2017gfo lasted between $\sim$ days, $\sim$ weeks, and $\sim$months depending on the observed wavelength (e.g., data compilation by \citealt{Villar17} and references therein; \citealt{Kasliwal22}).
%The modeling of the light curves and spectra of AT\,2017gfo has been used to extract information about the heavy-metal content (\textbf{typically parametrized by} the mass fraction of lanthanides, \Xlan, a product of $r$-process nucleosynthesis and subset of the heavy metals), as well as the mass and velocity of the different ejecta components. Accurate estimates of the mass, velocity, and \Xlan\, are crucial because the properties of kilonova ejecta can be used to understand the fate of the merger remnant (e.g., \citealt{Margalit17,Radice20,Radice23}), which constrains the neutron star equation of state, and constrains the contribution of kilonovae to the evolution of the heavy-metal content of the Universe (e.g., \citealt{Hotokezaka15,Qian&Wasserburg07,Wallner15,Ji16,Kasen17,Rosswog18}). 
\vspace{-0.1cm}

While the emission of AT\,2017gfo at $\gtrsim 1$ day is well-explained by the radioactive decay of $r$-process material (e.g., \citealt{Kasen17}), the source of early ($\lesssim$ 1 day) blue emission remains unclear (e.g., \citealt{Arcavi18}). There are a plethora of emission mechanisms that can operate at early times to affect early blue emission: free neutron decay \citep{Metzger15,Gottlieb20}, jets, shocks, and cocoon emission \citep{Gottlieb18a,Piro18, Banerjee26}, radioactive heating from heavy elements \citep{Banerjee20}, and central-engine energy injection \citep{Ai25} that can become difficult to distinguish on $\sim$ day timescales as the source fades and $r$-process heating dominates. Here, we combine the latest atomic data of highly ionized $r$-process elements \citep{Banerjee20,Banerjee22,Banerjee24} and quasi-NLTE techniques \citep{Brethauer26} to produce models of early-time blue emission from the decay of free neutrons, with specific focus on potential signals at $t\lesssim$ hours that can be detected by UV missions such as UVEX \citep{Kulkarni21}, ULTRASAT \citep{ULTRASAT}, and Habitable Worlds Observatory \citep{HabitableWorlds26}, as well as X-ray missions like Einstein Probe \citep{EinsteinProbe25} for the first time. 

General-relativistic magnetohydrodynamic (GRMHD) simulations have shown that during the merger process of a binary neutron star (BNS) system material can be ejected at high velocities ($\gtrsim 0.4c$) and possibly contain free neutrons (e.g., \citealt{Goriely14,Just15,Radice18b,Combi23,Schnabel26}). The free neutrons then $\beta$-decay, powering short-lived emission that peaks on $\sim$ hours timescale (e.g., \citealt{Metzger15,Gottlieb20}). 

While free neutrons are expected to be in the fastest moving dynamical ejecta (e.g., \citealt{Metzger15,Schnabel26}), the specific density profile of free neutrons is not well constrained, including to what maximum velocity the free neutrons extend. After a BNS merger, \cite{Ishii18} find that the mass fraction of neutrons becomes constant at a sufficiently high velocity by $\sim0.1 $s, though hydrodynamical effects may alter the final density profile. The total mass of free neutrons is somewhat more constrained; initial smooth particle hydrodynamics code simulations found $\sim 10^{-4}$ M$_\odot$ of fast-moving material ($\gtrsim 0.6c$) that could contain free neutrons \citep{Bauswein13,Goriely14,Just15,Metzger15}, though more recent grid-based hydrodynamics and higher resolution simulations estimate between 10$^{-7}$ and 10$^{-5}$ M$_\odot$ of fast-moving material \citep{Ishii18,Radice18}, with \cite{Dean21} and \cite{Schnabel26} resolving up to $\sim$ few 10$^{-5}$ M$_\odot$. If instead of a gravitational wave-driven inspiral, the neutron stars merge in a head-on collision (such as a highly eccentric orbit via dynamical mergers), then as much as $\sim$ 10$^{-2}$ M$_\odot$ high-velocity material can be ejected \citep{Dean21}.

The free neutrons inject additional energy into the KN ejecta via $\beta$-decay, resulting in a proton and a high-energy electron. Following \cite{Kulkarni05}, on average each free neutron emits a 302.77 keV electron (with the remaining $\sim500$ keV carried away by the neutrino) with a half-life of $\tau_n \approx$ 613.9 s \citep{Workman22}. The electron will then deposit its energy through Coulomb interactions with free electrons as well as via excitations and ionizations. Thus, the decaying neutrons can produce bright emission while simultaneously dominating the radioactive ionization rate at early times and ionize material to higher states, similar to the electrons from $\beta$-decay of $r$-process material \citep{Hotokezaka21,Pognan23,Brethauer26} which could change the spectral energy distribution significantly.

The emission from free neutrons has been modeled in a limited number of studies, either analytically or semi-analytically \citep{Metzger15,Ishii18,Gottlieb20,Dean21, Magistrelli24,Schnabel26}. We present here for the first time a detailed radiative transfer simulation study using frequency-dependent opacities from realistic atomic data, as well as incorporating non-thermal ionization effects.

The paper is organized as follows: in \S \ref{sec:methods} we discuss the radiative transfer code \texttt{Sedona} we employ for our models, the set up of our models, and underlying equations. We present the observable signatures in \S \ref{Sec:Observables}, then compare them to previous works and discuss complications to observing the free neutron signal in \S \ref{Sec:Discussion}. Finally, we conclude the paper in \S \ref{Sec:Conc}.

\section{Methods} \label{sec:methods}
\subsection{Sedona and Initial Conditions Setup} \label{sec:Sedona}

We use the time and wavelength-dependent Monte Carlo radiative transfer code \texttt{Sedona} \citep{Kasen06,Roth15} to generate synthetic spectra, from which we derive light curves. Each model is spherically symmetric and we explore the effects of free neutrons by simulating a grid of free neutron masses and density profiles added to a fiducial model (see Section \ref{subsec:SetupNeutron} for more details).

We run each simulation in 1D with 120 zones with physical conditions defined by a temperature, density, velocity, and chemical composition. 
\texttt{Sedona} subsequently evolves the system under the assumption of homologous expansion.
Neutron star merger ejecta generally approach homologous expansion on timescales of order $\sim$seconds \citep{Grossman14,Rosswog14,Sippens25}.  Small departures from homology, caused by radioactive heating and interactions
between different ejecta components, may persist longer; the ejecta become
highly homologous (deviations of $\sim1\%$) by timescales of minutes to hours \citep{Grossman14,Rosswog14,Sippens25}. 

We use an ejecta density profile of a broken power law
\begin{equation}
        \rho (v,t) =
        \Biggl\{ \begin{array}{ll}
            C_\rho \dfrac{M}{v_t^3t^3} \left(\frac{v}{v_t}\right)^{\delta} & v \leq v_t  \\
            \\
             C_\rho \dfrac{M}{v_t^3t^3} \left(\frac{v}{v_t}\right)^{q} & v_t < v < v_{\rm max}, 
        \end{array} 
\end{equation}
\noindent where $\delta$ and $q$ are the power law index of the inner and outer ejecta, respectively, $v_t$ is the velocity at which the transition between the two power law indices occurs, $t$ is time, $M$ is the total ejecta mass, \vmax\, is the maximum velocity the ejecta extends to, and $C_\rho$ is the normalization constant. Following \cite{Kasen17} we adopt the broken power-law density profile with $\delta = -1$ and $q = -10$, which is similar to accretion disk wind models (e.g., the velocity distributions studied in \citealt{Fryer24}) and supernova ejecta from a massive compact star such as a Wolf-Rayet (e.g., \citealt{Chevalier89}).
The transition velocity, $v_{t}$, is defined by
\begin{equation}
    v_t \equiv C_v v_k = C_v \sqrt{\frac{2E_k}{M}}, 
\end{equation}
\noindent where $C_v$ is the normalization constant to ensure the ejecta have total kinetic energy $E_k$. 

The composition of the fiducial dynamical ejecta is based on solar abundance patterns (or meteoric where solar abundances are not available) presented in \cite{Asplund09}, and $r$-process residuals from \cite{Simmerer04} for elements with atomic number Z = 31--70. The composition is then normalized by mass fraction such that all elements of Z = 58--70 have a total mass fraction of \Xlan \, and all other elements have a total mass fraction of 1-\Xlan. We do not consider any elements of Z $\geq$ 71. 

Each spectrum is calculated every $\frac{\delta t}{t}$ = 0.2 until 0.1\,d where spectra are taken every 0.1\,d with 2309 logarithmically-spaced frequency points every 0.005$\nu_0$ between $10^{13}$ and $1 \times 10^{18}$ Hz (30 \AA - 30 $\mu$m). We limit hydrodynamical time steps to the minimum of 1\% of the elapsed time and 0.2\,d, which is sufficient to resolve the expansion evolution of the ejecta. The simulation is initialized assuming LTE, and then switches to QNLTE \citep{Brethauer26} after the first hydrodynamical step. 

At each time step, Monte Carlo packets (effectively bundles of photons of a given wavelength that total up to a specified energy amount) are released in accordance to the radioactive/$r$-process heating rate convolved with the thermalization efficiency (see \citealt{Brethauer24,Brethauer26} for more details) and interact with a zone through scattering and absorption. Monte Carlo photons that reach the outer edge of the simulation (defined by the maximum velocity) escape the ejecta and are collected and binned in time and frequency to generate the spectral time series of the model, with all relevant Doppler shift, beaming, time dilation, and light travel-time effects taken into account for an observer infinitely far away.

\vspace{-0.5cm}
\subsection{Atomic Data}
All models use atomic data generated by the Hebrew University Lawrence Livermore Atomic Code (\texttt{HULLAC}, \citealt{Bar-Shalom01}) for elements Z = 26--88, up to Z$^{+10}$ species, in a self-consistent and systematic way for a large number of elements, as presented in \cite{Tanaka20} and \cite{Banerjee24}. For this work, we use elements Z = 31--70, with Sm (Z = 62) to represent all lanthanide species for ionization stages $\geq \rm{Z}^{+4}$ as the high velocity of the photosphere at early times ($\gtrsim 0.5c$) will result in highly blended features regardless, making individual lines less important. All opacities employ the Sobolev expansion opacity for binning the large number of lines from lanthanides. The Sobolev expansion opacity uses the Sobolev approximation \citep{Sobolev60} for bound-bound transitions, which is applicable when the thermal line width is negligible compared to that of the expansion velocity. This is true for KN ejecta, as the expansion velocity is typically of the order 10$^3$ km s$^{-1}$ while the thermal velocities are of the order of 1 km s$^{-1}$ \citep{Kasen13}, as well as early times where the photosphere is within high-velocity ($\gtrsim 0.5c$) and hot ($\sim10^5$ K) ejecta. The lines are then binned within the broader frequency bins of the simulation. We include bound-free (assuming hydrogen-like cross sections), free-free, and electron scattering opacities in each model. Prior to $\sim1$ hr, bound-bound transitions dominate the opacity at $\lesssim100$ eV with bound-free becoming dominant $\gtrsim100$ eV (see \S \ref{Subsec:Opac} for more details)

\vspace{-0.4cm}
\subsection{Free Neutron Models \label{subsec:SetupNeutron}}

\begin{figure}
    \centering
    \includegraphics[width=0.49\textwidth,trim={0cm 0cm 0cm 0cm},clip]{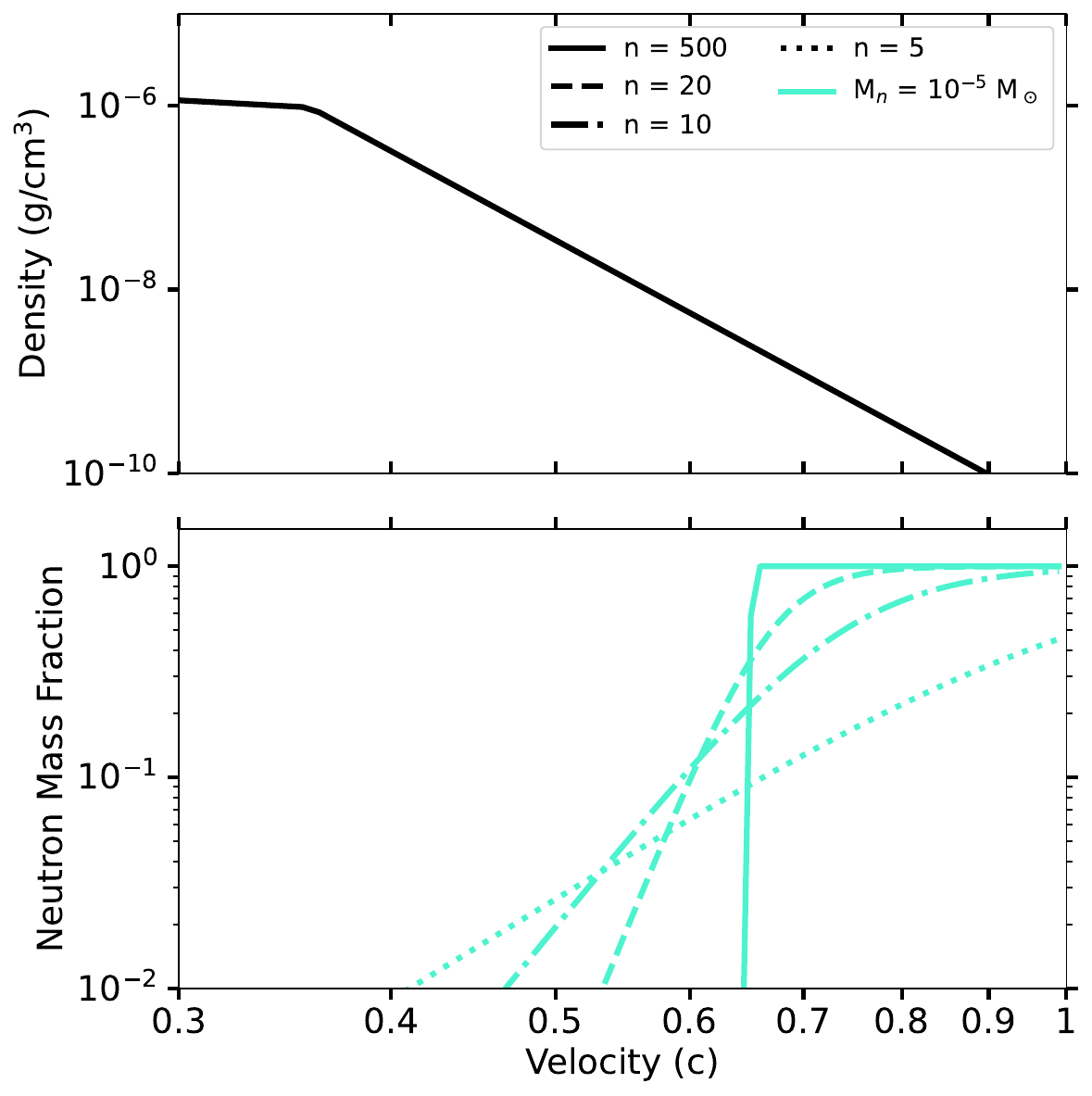}
    \caption{\textit{Top Panel:} Density profile at $t \sim 1.4$ minutes for M$_n$ = $10^{-5}$ M$_\odot$ embedded in the fiducial dynamical ejecta (blue) of varying neutron mixing  \textit{Bottom Panel:} Fraction of initial free neutron mass as a function of position in the ejecta. As $n$ increases, there is less mixing of the free neutrons with the ejecta and therefore less homogeneously mixed.}
    \label{Fig:NeutronRho}
\end{figure}

We vary the mass and the mixing profiles of the free neutrons embedded in a fiducial dynamical ejecta, with properties typical of dynamical ejecta from GRMHD simulations: mass M = 0.003 $M_\odot$, $v_k$ = 0.3$c$, and \Xlan = 10$^{-1}$ which corresponds to $Y_e \lesssim 0.2$ (e.g., \citealt{Bauswein13,Hotokezaka13,Lippuner15,Sekiguchi16,Ciolfi17,Radice18}). The grid of models includes points with free neutron masses of 10$^{-7}$, 10$^{-6}$, 10$^{-5}$, and 10$^{-4}$ M$_\odot$. We run an additional grid with an ejecta mass M = 0.03 $M_\odot$ to investigate $M_n$ = 10$^{-3}$ and 10$^{-2}$ M$_\odot$ as well (see Table \ref{tab:models}) as the effects of a dynamical ejecta mass more similar to that found in AT2017gfo (e.g., \citealt{Villar17,Rastinejad25}; though see complications of systematic errors in mass estimates in \citealt{Brethauer26}). While we include models that contain $10^{-3}$ and $10^{-2}$ M$_\odot$ of free neutrons, we emphasize that this would require an extreme scenario (see \citealt{Dean21} for more details). The free neutrons are placed by changing the composition of the ejecta according to the equation
\begin{equation}
    X_n(v) = X_{n, out} - X_{n, out} \bigg(\big(\frac{v}{v_{char}}\big)^n+1\bigg)^{-1}
\end{equation}
\noindent where $X_{n, out}$ is the maximum neutron mass fraction (set to 0.999, as \citealt{Schnabel26} shows the value approaches 1), $n$ is the steepness of the composition transition corresponding to the mixing of the free neutrons with the rest of the dynamical ejecta, and $v_{char}$ is the characteristic velocity of the transition and is determined by normalizing $M_n$. In our grid, we use $n = 5, 10, 20$ and 500 to represent a broad range of potential mixing, though we expect more stratified ejecta ($n = 500$) to be more realistic due to the nucleosynthesis conditions \citep{Schnabel26}. An example set of composition profiles for the M$_n$ = $10^{-5}$ M$_\odot$ models is shown in the bottom panel of Figure \ref{Fig:NeutronRho}. At smaller values of $n$, the free neutrons are less stratified in the high-velocity tail and the free neutrons are therefore more mixed with the dynamical ejecta. 

To demonstrate the effect of the choice of \vmax\,, we vary \vmax\, with values 0.6, 0.7, 0.8, 0.9, 0.95, and 0.99$c$. 

\begin{table}
    \centering
    \begin{tabular}{c|cc}
        \hline
        \hline
        M$_n$ ($M_\odot$) & $n$ & $M$ ($M_\odot$) \\
        \hline
        10$^{-7}$ & All& $3\times10^{-3}$ \\
        10$^{-6}$ & All& $3\times10^{-3}$\\
        10$^{-5}$ & All& $3\times10^{-3}$\\
        10$^{-4}$ & All& $3\times10^{-3}$\\
        10$^{-3}$ & All& $3\times10^{-2}$\\
        10$^{-2}$ & 500& $3\times10^{-2}$\\
    \end{tabular}
    \caption{Models parameters used in our grid of simulations. Simulations with M$_n$ < $10^{-3}$ use a dynamical ejecta mass of $3\times10^{-3}$ M$_\odot$, while the remainder use $3\times10^{-2}$ M$_\odot$. We vary the value of $n$ between 5, 10, 20, and 500 for all models except M$_n$ = $10^{-2}$ M$_\odot$ to prevent neutrons from mixing far into the bulk of the ejecta where it would be unphysical (e.g., \citealt{Schnabel26}) All models vary \vmax\, between 0.6, 0.7, 0.8, 0.9, 0.95, and 0.99$c$.}
    \label{tab:models}
\end{table}

\vspace{-0.6cm}

\section{Observables} \label{Sec:Observables}

\begin{figure}
    \centering
    \includegraphics[width=0.47\textwidth,trim={0.7cm 0.6cm 0cm 1.8cm},clip]{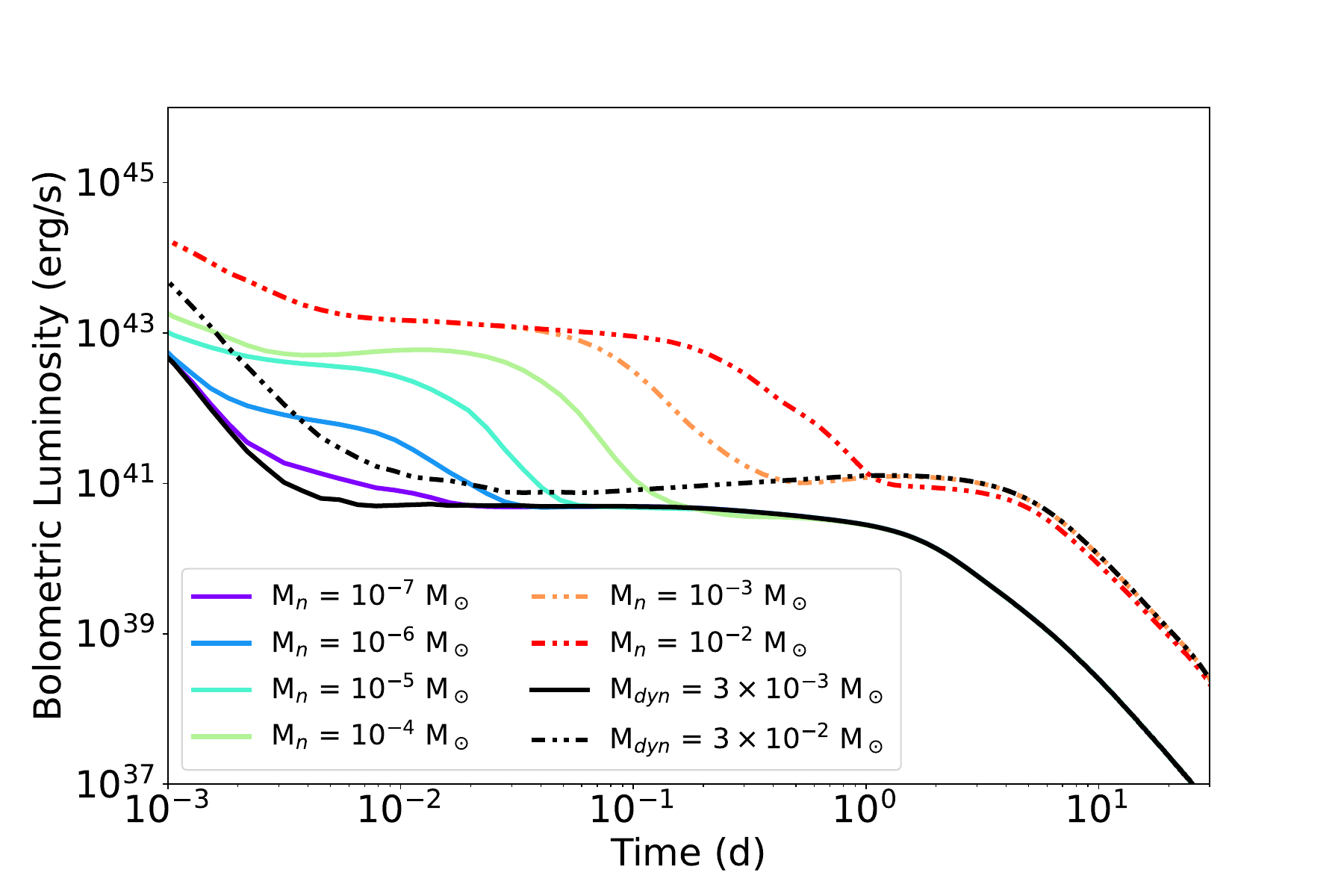}
    \caption{Bolometric luminosity curves for varying neutron masses with $n = 500$ composition profile extending to \vmax\,= 0.7$c$. In black is a dynamical ejecta model with no free neutrons embedded within the ejecta.}
    \label{Fig:ConstBol}
\end{figure}

\begin{figure*}
    \centering
    \includegraphics[width=0.97\textwidth,trim={0cm 0cm 0cm 0cm},clip]{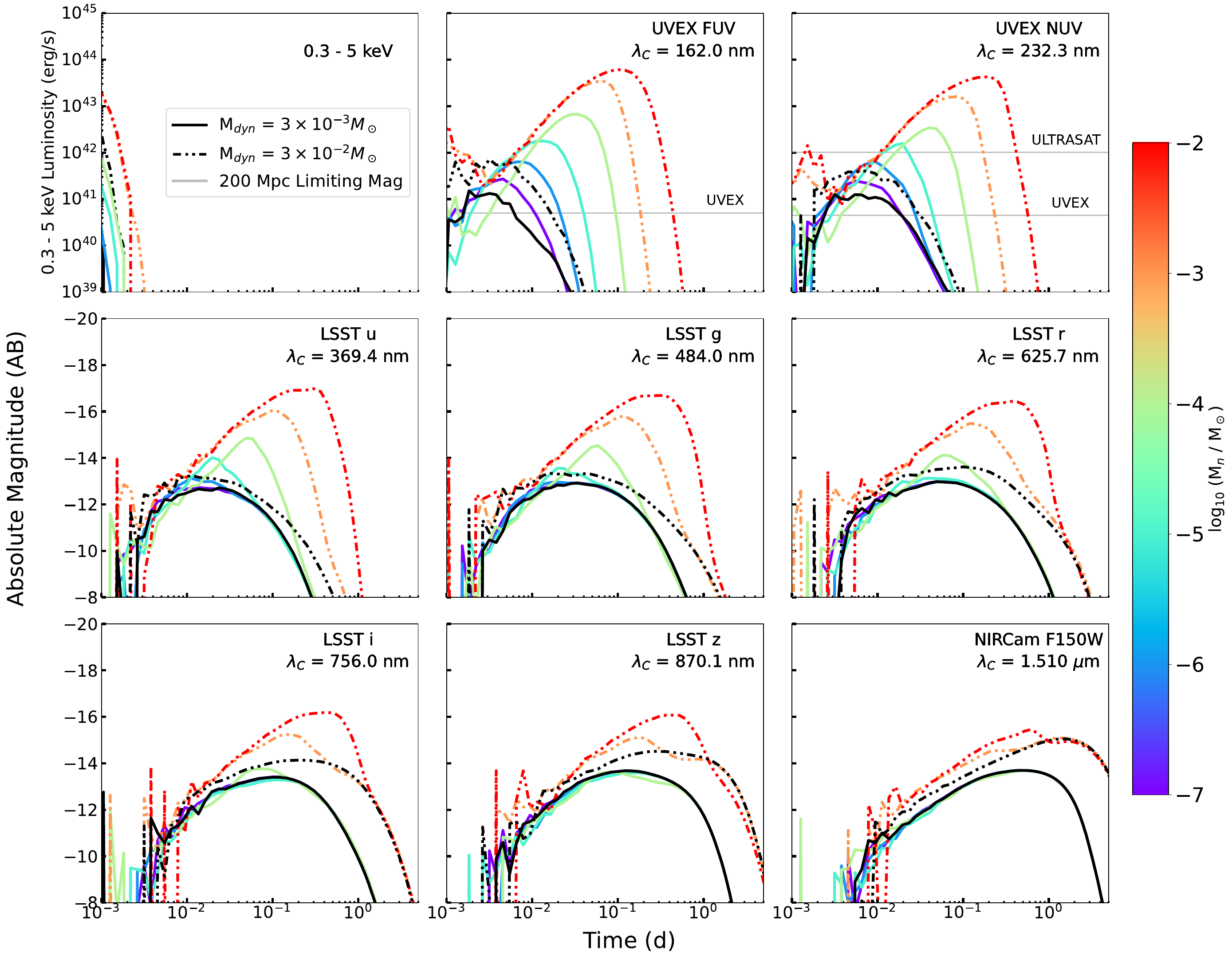}
    \caption{Light curves for varying neutron masses (colors) with $n = 500$ composition profile of free neutrons with $v_{\rm max} = 0.7c$. The typical depth of UVEX and ULTRASAT photometric observations at 200 Mpc are shown in gray. In black is a $3\times10^{-3}$ and $3\times10^{-2}$ M$_\odot$ dynamical ejecta model with no free neutrons (solid and dash-dot-dotted, respectively). The UV flash peaks on typical timescales of $\sim5$ minutes to $\sim2$ hours, depending on M$_n$. Neutron-free models and neutron-rich models are most distinguishable in the FUV due to the distinct peak at FUV wavelengths. Monte Carlo noise is most prominent at early times and at redder wavelengths due to the short time bin width and weak red emission.
    %Additionally, the presence of free neutrons can also be seen indirectly through the fainter and more rapidly decaying post-peak light curves at UV and optical wavelengths in lower neutron mass models compared to the neutron-free ejecta caused by enhanced UV/optical opacities at those times. With QNLTE, the outermost layers of the neutron-free ejecta are ionized to triply ionized species due to the low densities. However, when there are free neutrons in the ejecta, the hydrogen generated from $\beta$-decay absorbs some of the energy that went into ionizing the $r$-process material, resulting in less ionization per ion and thus lowering the typical ionization state and leading to higher UV/optical opacities.
    }
    \label{Fig:ConstLC}
\end{figure*}

\begin{figure}
    \centering
    \includegraphics[width=0.49\textwidth,trim={0cm 3cm 0cm 4.5cm},clip]{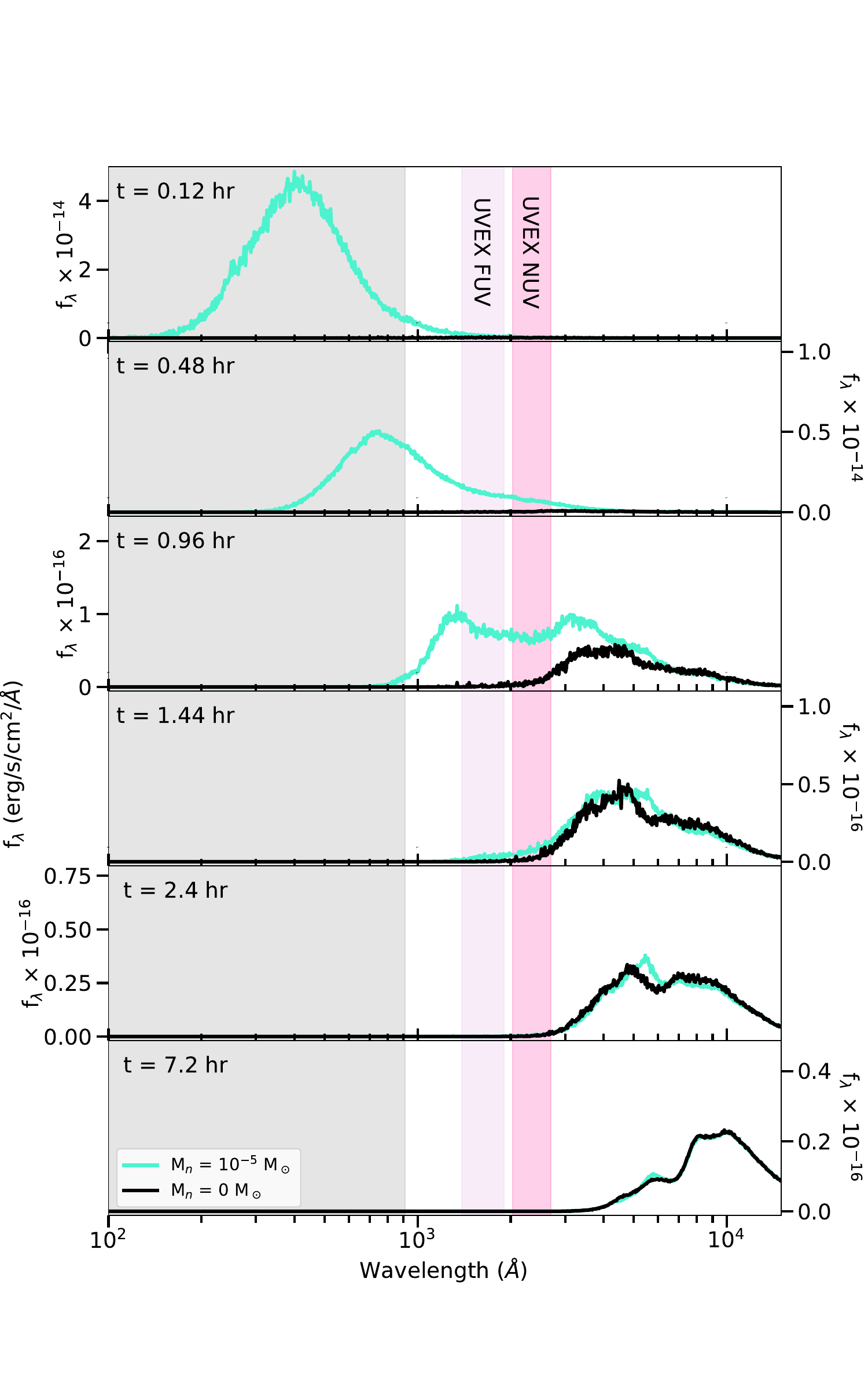}
    \caption{Spectral sequence for $n = 500$, $v_{\rm max} = 0.7c$, and $M_n = 10^{-5}$ M$_\odot$ model with UVEX FUV and NUV filter regions highlighted at an assumed distance of 40 Mpc compared to a neutron-free model. Region of high galactic absorption due to hydrogen is highlighted in gray. The SED rapidly shifts from peaking in the EUV to the optical and IR on $\sim$hrs timescale.}
    \label{Fig:SpecSeq}
\end{figure}

\begin{figure*}
    \centering
    \includegraphics[width=0.97\textwidth,trim={0cm 0cm 0cm 0cm},clip]{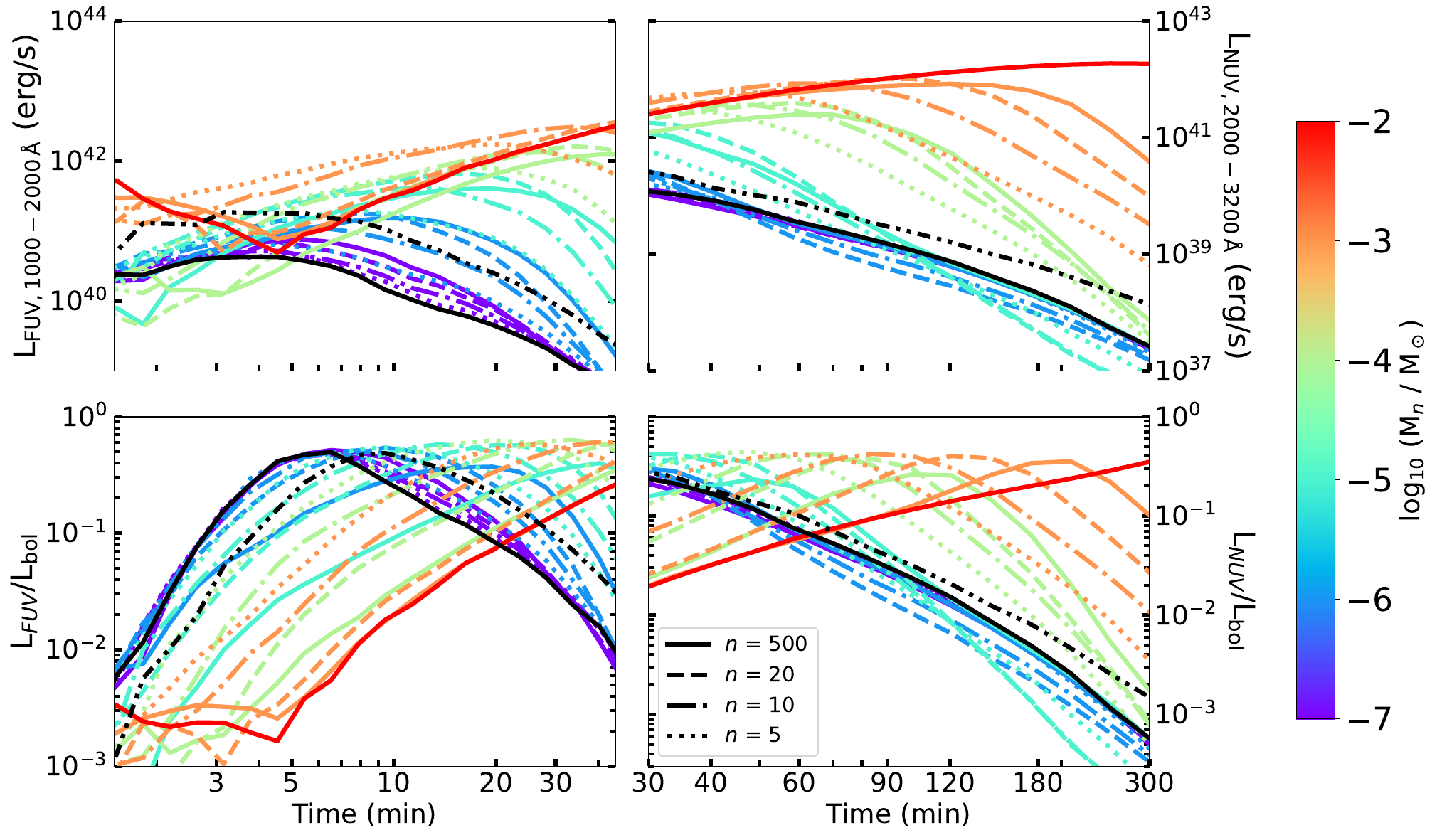}
    \caption{\textit{Top Left Panel:} Zoom-in on FUV luminosity from Figure \ref{Fig:ConstLC}, with varying composition profile of free neutrons (linestyle) from minimally mixed ($n = 500$) to highly mixed ($n = 5$) for $v_{\rm max} = 0.7c$ models. The neutron-free models are shown in black. For $M = 3\times10^{-3}$ M$_\odot$ models, higher degrees of mixing of the free neutrons with the dynamical ejecta result in suppression of FUV emission, as the longer diffusion timescale reprocesses emission to optical emission at later times. \textit{Bottom Left Panel:} Fraction of bolometric luminosity emerging as FUV, with each model peaking at $\sim60\%$ and rapidly declining post-peak. \textit{Top Right Panel:} Same as Top Left Panel, but for NUV luminosity. For free neutron mass $\lesssim 10^{-5}$ M$_\odot$, the additional hydrogen atoms leads to less efficient ionization in $r$-process elements which absorb NUV radiation and cause the transient to appear dimmer in the UV compared to the neutron-free model. \textit{Bottom Right Panel:} Same as Bottom Left Panel, but for NUV luminosity. In the first hour, most of the luminosity that escapes is in the FUV/NUV indicating the importance of UV coverage to get accurate bolometric luminosities.}
    \label{Fig:XUVLC}
\end{figure*}

%\subsection{Central Engine \label{Subsec:Central}}

\begin{figure*}
    \centering

    \includegraphics[width=0.97\textwidth,trim={0cm 0cm 0cm 0cm},clip]{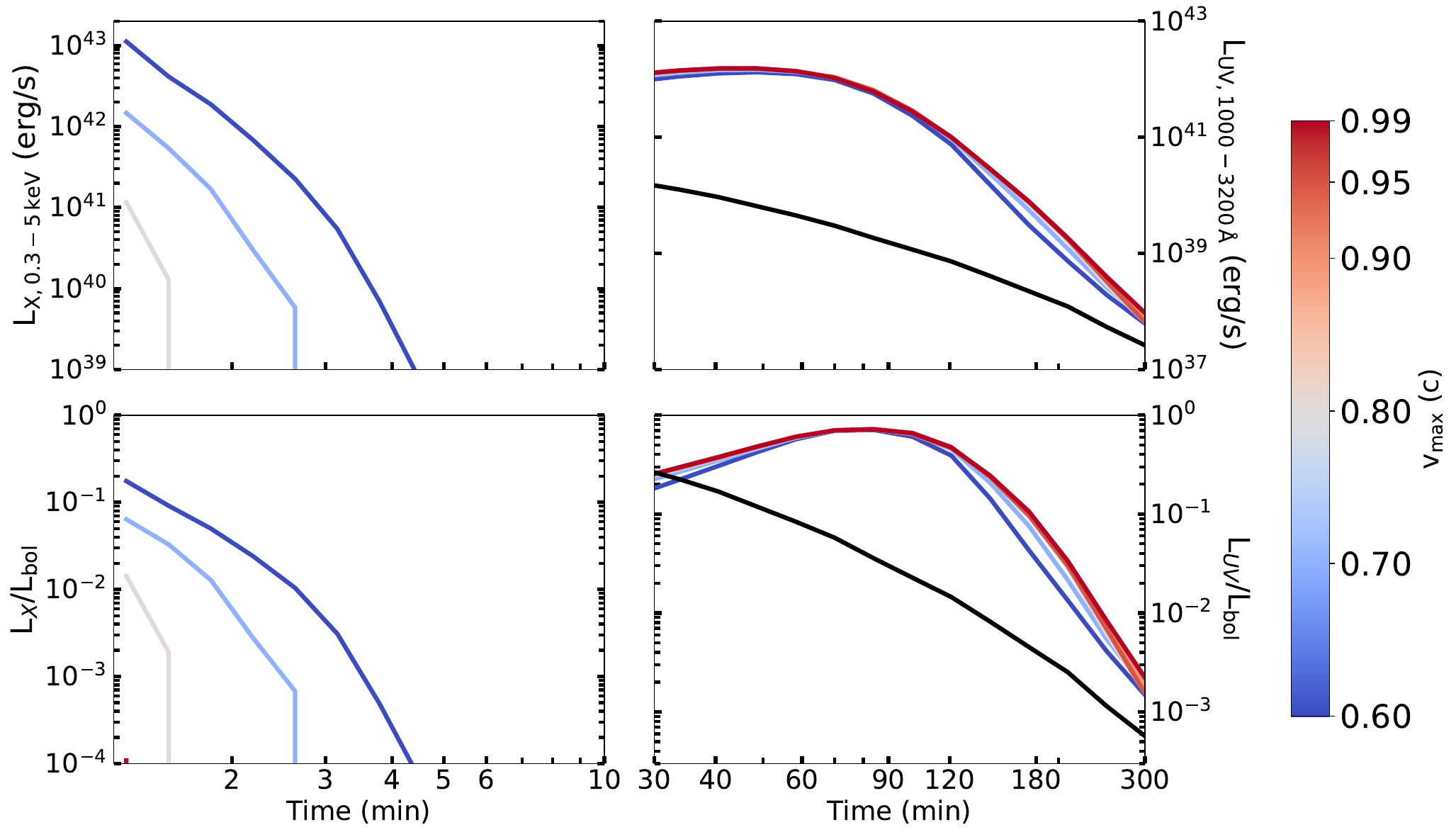}
    \caption{Same as Figure \ref{Fig:XUVLC}, except varying the maximum velocity that the free neutrons extend to for a M$_n = 10^{-4}$ M$_\odot$ model with $n = 500$. In black is the neutron-free dynamical ejecta model that extends to \vmax\, = 0.9c. As the maximum velocity is decreased, blue emission is less reprocessed and the photospheric radius is smaller, resulting in higher temperatures and therefore a higher fraction of emission emerges as brighter X-rays flashes, reaching $\sim10\%$ at $v_{\rm max} = 0.6c$ while $v_{\rm max} \geq 0.9c$ do not produce enough X-rays to be simulated.}
    \label{Fig:MaxVelLC}
\end{figure*}

\begin{figure*}
    \centering
    \begin{subfigure}[b]{0.49\textwidth}
         \centering
         \includegraphics[width=\textwidth,trim={0cm 3cm 0cm 4.5cm},clip]{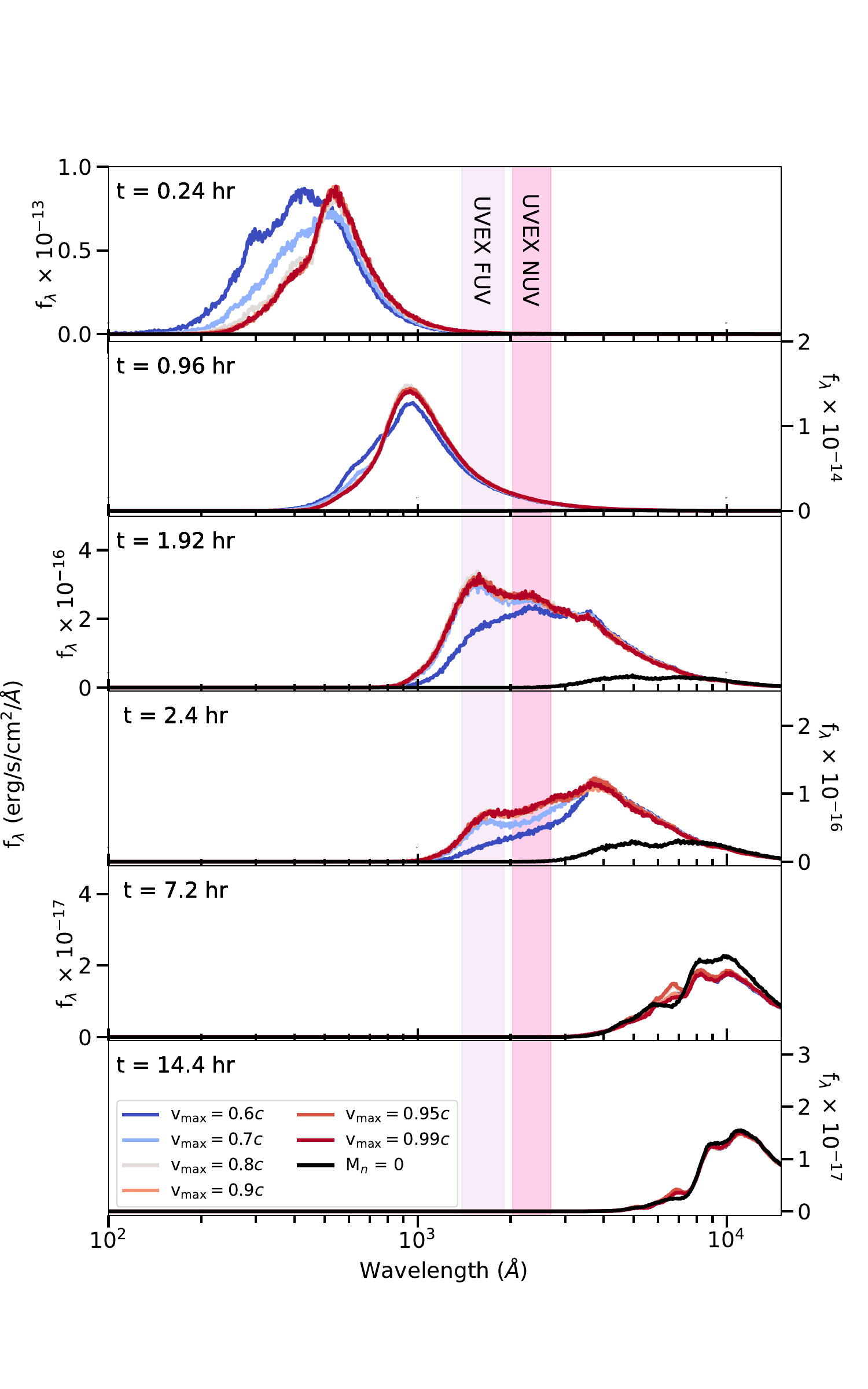}
         
    \end{subfigure}
    \begin{subfigure}
        [b]{0.49\textwidth}
         \centering
         \includegraphics[width=\textwidth,trim={0cm 3cm 0cm 4.5cm},clip]{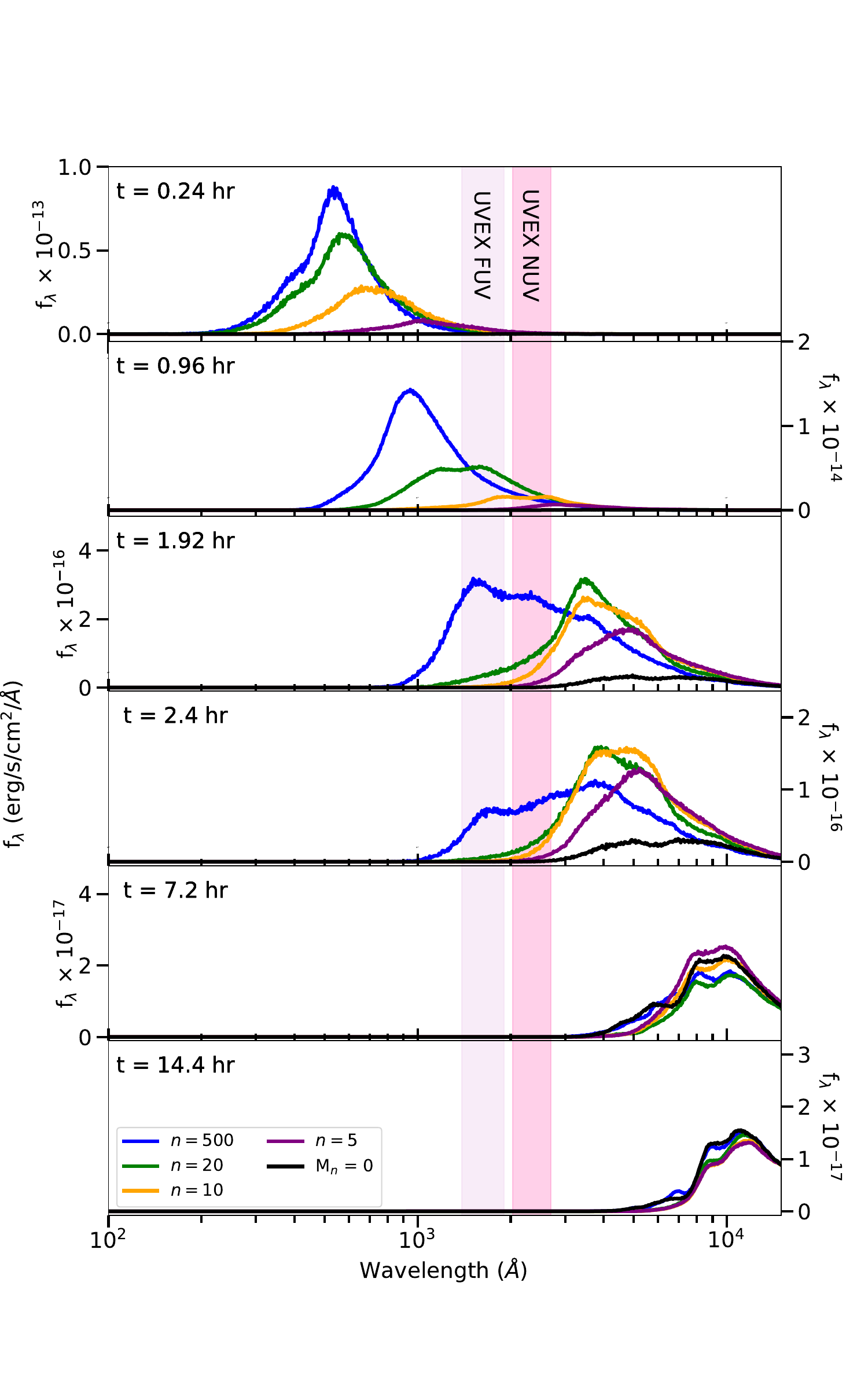}

    \end{subfigure} 
    \caption{\textit{Left Panel:} Spectral sequence for $n = 500$, $M_n = 10^{-4}$ M$_\odot$ and variable \vmax\, with UVEX FUV and NUV filter regions highlighted at an assumed distance of 40 Mpc. The peak wavelength and peak flux of the spectrum shift in response to the varying temperature and photospheric radius at $t \lesssim 1$ hr, with lower \vmax\, resulting in hotter but smaller photosphere. \textit{Right Panel:} Same as left panel, but for variable values of $n$ and \vmax\, is held constant at 0.9$c$. As the neutrons are more mixed with the dynamical ejecta, the spectral peak at $t \lesssim$ hours shifts redward and becomes fainter, resulting in a delayed and brighter red peak due to higher opacities.}
    \label{Fig:SpecSeqDensVel}
\end{figure*}

\subsection{Bolometric Luminosity, Light Curves, and Spectra}

At free-neutron masses $\gtrsim10^{-6}$ M$_\odot$, the early-time bolometric luminosity is dominated by the emission of free neutrons, as shown in Figure \ref{Fig:ConstBol}. The $n = 500$ neutron-rich models demonstrate enhanced plateau-like luminosity over the neutron-free model (shown in black) to times that scale with the free neutron mass. The plateau qualities are likely a result of optical depth effects with the photosphere receding through almost exclusively hydrogen ejecta for less mixed ejecta with time until reaching the bulk of $r$-process material akin to a Type IIP supernova as the hydrogen-rich envelope remains ionized. Afterwards, the bolometric luminosity relaxes to that of the underlying dynamical ejecta emission as the free neutron emission has entirely escaped, the ejecta cool, and the high-velocity ejecta become optically thin. Only at the most massive models is the late-time bolometric luminosity fainter than the neutron-free case due to a significant fraction of the ejecta being converted into free neutrons that no longer heat the ejecta according to the r-process heating rate. The light curves and spectral sequence shown in Figures \ref{Fig:ConstLC} and \ref{Fig:SpecSeq} for the same models reveal that much of the enhanced emission emerges at Extreme Ultraviolet (EUV, $\sim$125--1000 \AA), Far Ultraviolet (FUV, $\sim$1000--2000 \AA), and Near Ultraviolet (NUV, $\sim$2000--3200 \AA) wavelengths during this time. The different free neutron masses are most distinguishable both photometrically and spectroscopically at EUV and FUV wavelengths, with even the $M_n = 10^{-7}$ M$_\odot$ model showing a distinct FUV excess over the bare dynamical ejecta at $t \lesssim 30$ minutes. However, when the free neutrons are more mixed with the dynamical ejecta, the UV emission is suppressed and reprocessed to longer wavelengths that results in brighter optical emission making it more difficult to distinguish the $M_n = 10^{-7}$ M$_\odot$ model from the neutron-free model (Figure \ref{Fig:XUVLC}). Similarly, \vmax\, determines the amount of material that emission travels through and how much the emission gets reprocessed to longer wavelengths. Figure \ref{Fig:MaxVelLC} shows that at higher \vmax, the X-ray emission at early times is more reprocessed to UV emission at later times. 

At $t \lesssim 2$ hrs, the spectra can show deviations from a blackbody, such as exhibiting multiple peaks (Figures \ref{Fig:SpecSeq}, \ref{Fig:SpecSeqDensVel}). Despite the high velocities of the photosphere, the presence of features that vary with $M_n$ and \vmax\, suggest that spectroscopic observations at early phases can grant further insight over photometric observations.

The presence of free neutrons can also counterintuitively lead to a \textit{decrease or more rapid decline} in UV/optical light curves compared to neutron-free models post-UV peak for more highly mixed models due to changes in opacity. Broadly speaking, there are two competing effects that can change the opacity as the ejecta become increasingly dominated by free neutrons. As the free neutron fraction increases, the resulting hydrogen generally remains ionized out to $\sim$days and the dominant opacity source becomes electron scattering with $\kappa_{es} \sim 0.4$ cm$^2$ g$^{-1}$, reducing the opacity. However, for a given ejecta mass, a higher neutron mass fraction both reduces the amount of $r$-process material that is decaying and ionizing the surrounding ejecta while simultaneously increasing the number density of atoms. Thus, the average ionization energy per ion decreases, producing a lower ionization state and higher opacity from the remaining $r$-process material. In the regime where the opacity in the outermost layers grows faster from the lower ionization state than the dominance of electron scattering can reduce the opacity, the light curve will be fainter or show a more rapid decline than the neutron-free ejecta. Eventually, the ejecta expand and the outermost layers become optically thin and the effect vanishes. The additional reprocessing due to weakened ionization of $r$-process material is most prominent in Figure \ref{Fig:SpecSeqDensVel}, where some spectra at 7.2 hr show a significant redward shift compared to the neutron-free model for neutron-rich models. Interestingly, in the mixed models this feature is most observable at optical bands at $t\sim0.1-1$ days, even for neutron masses as small as $10^{-6}$ M$_\odot$, and could represent a unique signature of free neutrons if the UV emission is missed.
\vspace{-0.5cm}
\subsection{X-ray Flash}

Due to the short half-life of the free neutrons, the radioactive energy is deposited rapidly compared to expansion and $r$-process decay timescales, resulting in high temperatures while the ejecta is optically thick. The emission can be approximated as a blackbody with a temperature given by the following equation:
\begin{multline}
    T_{\rm eff} = \bigg(\frac{L_{\rm bol}}{4\pi \sigma R_{\rm ph}^2} \bigg)^{1/4} \approx 3.1\times10^5 \rm{K}\bigg(\frac{L}{3\times10^{43} \rm{erg\, s^{-1}}} \bigg)^{1/4} \\ \times \bigg( \frac{v}{0.8c}\bigg)^{-1/2} \bigg( \frac{t}{1.5 \,\rm{min}}\bigg)^{-1/2}
\end{multline}

\noindent which is sufficiently hot that the Wien tail of the blackbody can produce significant, potentially observable X-ray emission.

The escaping X-rays are generated by thermal emission. While thermal X-rays are commonly generated by the inner ejecta due to high temperatures, the X-rays are unable to escape as they are reprocessed to longer wavelengths within the high opacity lanthanides. However, the free neutrons are present in the highest velocity ejecta and therefore above much of the reprocessing lanthanides with electron scattering as the dominant opacity source. Thus, as the free neutrons decay and heat the ejecta to sufficiently high temperatures to produce X-rays, the conditions can be right for the X-rays to successfully escape. 

The strength of the X-ray flash is highly sensitive to the maximum velocity the ejecta extends to, \vmax, as well as the power-law index of the free neutron density profile; for $M_n = 10^{-4}$ M$_\odot$, the X-ray luminosity peaks between $\sim2\times10^{40}$ and 10$^{43}$ erg s$^{-1}$, comprising $<$0.1--$10\%$ of the total bolometric output at soft energy bands (Figure \ref{Fig:MaxVelLC}). At the smallest \vmax\, of 0.6$c$, the higher neutron density leads to increased heating and less absorption to produce a $\sim10^{43}$ erg s$^{-1}$ X-ray flash. However, at \vmax\, = 0.6$c$ the underlying dynamical ejecta is sufficiently compact that without neutrons the ejecta can achieve a $\sim 10^{42}$ erg s$^{-1}$ X-ray flash. The left side of Figure \ref{Fig:SpecSeqDensVel} demonstrates that smaller \vmax\, results in hotter blackbody temperatures but a smaller photospheric radius, leading to a fainter, bluer spectral peak that can more easily achieve thermal X-ray emission. Conversely, the right side of Figure \ref{Fig:SpecSeqDensVel} illustrates that higher degrees of mixing lead to a redder, fainter spectral peak that is less capable of producing X-rays.

The sensitivity to initial conditions produces a wide range in potential X-ray emission. At the most extreme, the $M_n = 10^{-2}$ M$_\odot$, $n = 500$, \vmax\, = 0.6$c$ model is capable of producing a L$_X\sim3\times10^{44}$ erg s$^{-1}$ flash, though this scenario may not be astrophysically probable due to the requirement for both a head-on collision of neutron stars and the fast-tail ejecta to be highly compressed. On the other hand, many (though not all) of the models with \vmax $\gtrsim 0.9c$ or $M_n \lesssim 10^{-5}$ M$_\odot$ do not produce detectable X-ray emission in our simulations.

Regardless of the density profile and \vmax, a crucial diagnostic of the free neutron X-ray flash is that the emission is spectrally soft as a result of arising from thermal emission. While the free neutron X-ray flash lasts on a similar timescale as the cooling emission from a cocoon, the cocoon emission is much harder \citep{Gottlieb20}. Additionally, the free neutron X-ray flash will be distinct from embedded central engine emission such as a magnetar, as the X-rays of a central engine could shine through the ejecta out to timescales of $\sim$ days \citep{Ai25}.

We end the section by commenting on the observability prospects of this X-ray signal. Detecting the X-ray flash will require instruments with a similar field of view to the Einstein Probe Wide-field X-ray Telescope (WXT) to detect due to the short duration of the X-ray emission ($\lesssim 5$ minutes, Figure \ref{Fig:ConstLC}). Assuming a typical 0.3--5 keV luminosity of $\sim3\times10^{42}$ erg~s$^{-1}$, the horizon distance for WXT to observe the X-ray flash from a binary neutron star merger would be $\sim 16$ Mpc for a flux limit of $10^{-10}$ erg~s$^{-1}$~cm$^{-2}$ for a 100 second exposure \citep{EinsteinProbe25}. If instead the merger was a head-on collision that produced M$_n$ $\sim10^{-2}$ M$_\odot$ and a L$_X \sim 3\times10^{44}$ erg~s$^{-1}$ X-ray flash, Einstein Probe could potentially observe the X-rays out to $\sim 160$ Mpc.  To calculate the rate of detection of such events, we assume the binary neutron star merger rate to be
%360 - 1800 events Gpc$^{-3}$ yr$^{-1}$ based on Chandra and XMM-Newton short gamma ray bursts \citep{RoucoEscorial23} and 
5 - 155  events Gpc$^{-3}$ yr$^{-1}$ based on a range of models presented in \cite{LIGO26}. Convolving the event rate of 5 (155) events Gpc$^{-3}$ yr$^{-1}$ with the field of view of WXT and assuming that all mergers produce a L$_X \sim 3\times10^{42}$ erg s$^{-1}$ X-ray flash from free neutrons, we expect WXT to observe one X-ray flash from a binary neutron star merger every $\sim138,000$ ($\sim 4,450$) years. If we instead consider a future instrument of similar field of view as WXT, in order to achieve a detection rate of 1 X-ray flash from a BNS merger per year, a limiting flux of $\sim 4\times10^{-14}$ ($\sim 4\times10^{-13}$) erg s$^{-1}$ cm$^{-2}$ would be required for a merger rate of 5 (155) events Gpc$^{-3}$ yr$^{-1}$. However, wide field instruments are not the only viable path forward. A small instrument with a $\sim$1 deg$^2$ field of view such as the proposed mission STAR-X \citep{StarX} and sufficient limiting flux would likely prove more fruitful; assuming a limiting flux of $10^{-15}$ erg s$^{-1}$ cm$^{-2}$ (approximately a factor of 3 more sensitive than Chandra, \citealt{Polzin23}) for such an instrument with a L$_X\sim3\times10^{42}$ erg s$^{-1}$ source would result in an expected observed X-ray flash rate of once per $\sim15.7$ years ($\sim6$ months) for a merger rate of 5 (155) events Gpc$^{-3}$ yr$^{-1}$.
\vspace{-0.4cm}
\subsection{UV Emission}
\label{Subsec:ObsUV}

The NUV/FUV peak produced by the free neutrons occurs on a timescale of $\sim$5 minutes to $\sim$ hours, depending on the mass and density profile of the free neutrons (Figure \ref{Fig:XUVLC}) with NUV peak magnitudes between $-13$ (M$_n\sim 10^{-7}$ M$_\odot$) and $-18$ (M$_n \sim 10^{-2}$ M$_\odot$) AB magnitude. The FUV behaves very similarly to the NUV except for the smallest free neutron mass which is fainter and fades more rapidly. 

The FUV and NUV are critically important for accurately constraining the bolometric luminosity. 40--60\% of the energy emerges at FUV/NUV wavelengths during the first few hours, and for free neutron masses $\lesssim10^{-4}$ M$_\odot$, the FUV/NUV observations can capture the peak of the spectral energy distribution (Figures \ref{Fig:SpecSeq}, \ref{Fig:XUVLC}). The FUV and NUV can also help to constrain the maximum velocity that the ejecta extend to, with higher \vmax\, models producing brighter FUV/NUV (Figure \ref{Fig:SpecSeqDensVel}). GRMHD simulations have so far had difficulty achieving sufficiently high resolution to accurately model the extent of the ejecta due to the small amount of mass contained at such high velocities, thus providing a feasible path forward to instead constrain \vmax\, observationally.

The effect of variable \vmax\, can be distinguishable from the effects of mixing. As $n$ decreases, the free neutrons are more thoroughly mixed with the dynamical ejecta. This leads to longer diffusion timescales, and flux is pushed to later times and redder wavelengths as interactions with lanthanides reprocess emission (Figure \ref{Fig:XUVLC}) and a fainter, redder spectral peak. The variation in peak magnitude as a result of variable mixing decreases with longer wavelengths, ranging from $\sim1$ mag fainter at UV wavelengths to negligible differences at NIR wavelengths across the mixing range simulated. Crucially, this is the converse of a larger \vmax, where the emission instead gets redder and potentially \textit{brighter} as opposed to fainter as with higher degrees of mixing.

If the peak of the UV emission is missed, rapid fading in UV bands ($\gtrsim 10$ mag day$^{-1}$) may be detectable and are characteristic of the presence of free neutrons. 

We end the section by commenting on the observability prospects of the UV signal. Detecting the free-neutron UV flash will require a telescope with rapid slewing and a wide field of view, such as the upcoming UVEX and ULTRASAT missions. For M$_n \sim 10^{-5}$ M$_\odot$ with a peak UV magnitude of $-15$ and a conservative NUV limiting magnitude of 25 for UVEX \citep{Singer25}, the photometric detectability horizon for observing the free neutron emission is $\sim 1$ Gpc when assuming no dust extinction. However, the average 3 hr response time to schedule Target of Opportunity events \citep{Kulkarni21} would result in getting on target well after the peak of the FUV/NUV emission from free neutrons ($t \sim 30-60$ minutes). The delay time may be prohibitive for follow-up, and so serendipitous detections with missions that have a larger field of view or faster response times may be more helpful. ULTRASAT has a similar NUV filter to UVEX and a shallower limiting magnitude 22.5 for a 900s exposure, but much larger field of view at $\sim204$ degrees$^2$ compared to the $\sim12$ degrees$^2$ of UVEX as well as an expected 15 minute response time \citep{ULTRASAT,Singer25}. At a NUV peak magnitude of -15, ULTRASAT will be capable of observing free neutron emission out to $\sim320$ Mpc. We similarly calculate the rate of serendipitous detection by convolving the BNS merger event rate of 5 (155) events Gpc$^{-3}$ yr$^{-1}$ \citep{LIGO26} with the field of view of each instrument and assume all mergers produce significant free neutrons to get an estimated rate of observing a free neutron UV flash every $\sim160$ ($\sim 5.2)$ years for UVEX and $\sim310$ ($\sim 9.8)$ years for ULTRASAT. To reach a detection rate of 1 free neutron UV flash per year, a future mission with a UVEX-like field of view would need an NUV limiting magnitude of $\sim28.7$ ($\sim26.2$). For an ULTRASAT-like field of view, the limiting magnitudes would have to reach $\sim26.6$ ($\sim24.1$).

\vspace{-0.4cm}
\subsection{Possibility of H-alpha Emission}

After the free neutrons decay and are converted into electrons and protons, the particles will continue to recombine and ionize leading to a potentially detectable signal from H$_\alpha$ emission if the recombination timescale is sufficiently short. 

The H$_\alpha$ emission will depend on the recombination coefficient defined by the equation from \cite{Axelrod80} for an optically thin medium:

\begin{equation}
    \alpha_{\rm rec,H} = 3\times10^{-13} \bigg(\frac{T}{10^4 {\rm K}}\bigg)^{-3/4} \rm{cm^3}s^{-1}
    \label{Eq:HRecomb}
\end{equation}

\noindent where $T$ is the gas temperature. We choose the optically thin medium to understand if the emission could be observable at late times, well after the free neutrons have decayed and when the ejecta are dilute. 

We estimate the potential H$_\alpha$ luminosity through the relation to the recombination rate through the equation

\begin{multline}
    L_{H\alpha} = h\nu_{\rm H\alpha} f N_Hn_e\alpha_{\rm rec} \\  \approx 6.9\times10^{33}\bigg(\frac{f}{0.45}\bigg) \bigg(\frac{n_e}{5\times10^4 \rm{cm ^{-3}}}\bigg) \bigg( \frac{M_n}{10^{-4} \rm{M}_\odot}\bigg) \bigg(\frac{T}{2500 \rm{K}}\bigg)^{-3/4} \rm{erg\, s^{-1}}
\end{multline}

where $f$ is the fraction of recombinations that cascade through the $n=3$ to $n=2$ states, $n_e$ is the electron number density, and $N_H$ is the number of hydrogen atoms. The densities of the relevant ejecta are too small to produce meaningful H$_\alpha$ emission by $t\sim5$ days, as the recombination timescale grows to $\sim$0.75 years for hydrogen in the fastest moving ejecta leading to ionization state freeze out. Detectable H$_\alpha$ emission would have to be produced at early times when densities are sufficiently high, as \cite{Perego22} found. \cite{Perego22} found that placing $\geq 10^{-5}$ M$_\odot$ of H on top of dynamical ejecta at $v$ = 0.33c with a luminosity $\lesssim 2\times10^{41}$ erg s$^{-1}$ was capable of producing hydrogen features in the KN spectra when directly calculating H level populations instead of assuming LTE between 0.2--0.3 days post merger. However, our models do not show signs of detectable H$_\alpha$ at these epochs, likely due to the combination of ionization state in the low-density ejecta from low recombination rates \citep{Brethauer26}, the high-velocity ($\gtrsim 0.5c)$ of the hydrogen that Doppler broadens any features, and the use of line expansion opacity which can underpredict the strength of individual strong lines.

\vspace{-0.6cm}
\section{Discussion} \label{Sec:Discussion}

\subsection{Comparison to Analytical Models}

The formalism provided by \cite{Metzger15} can be adapted to our models. The main difference between the models used in \cite{Metzger15} and this work is the distribution of free neutrons within the ejecta; \cite{Metzger15} used a distribution proportional to $\arctan(m_n/m)$ \footnote{Equation 7 in \cite{Metzger15} was mistakenly written as $\arctan(m/m_n)$} where $m$ is the mass coordinate of the ejecta and $m_n$ is the transition mass, where the ejecta becomes dominated by free neutrons.  

\begin{figure}
    \centering
    \includegraphics[width=0.99\linewidth,trim={2.6cm 2.4cm 1.2cm 2.4cm},clip]{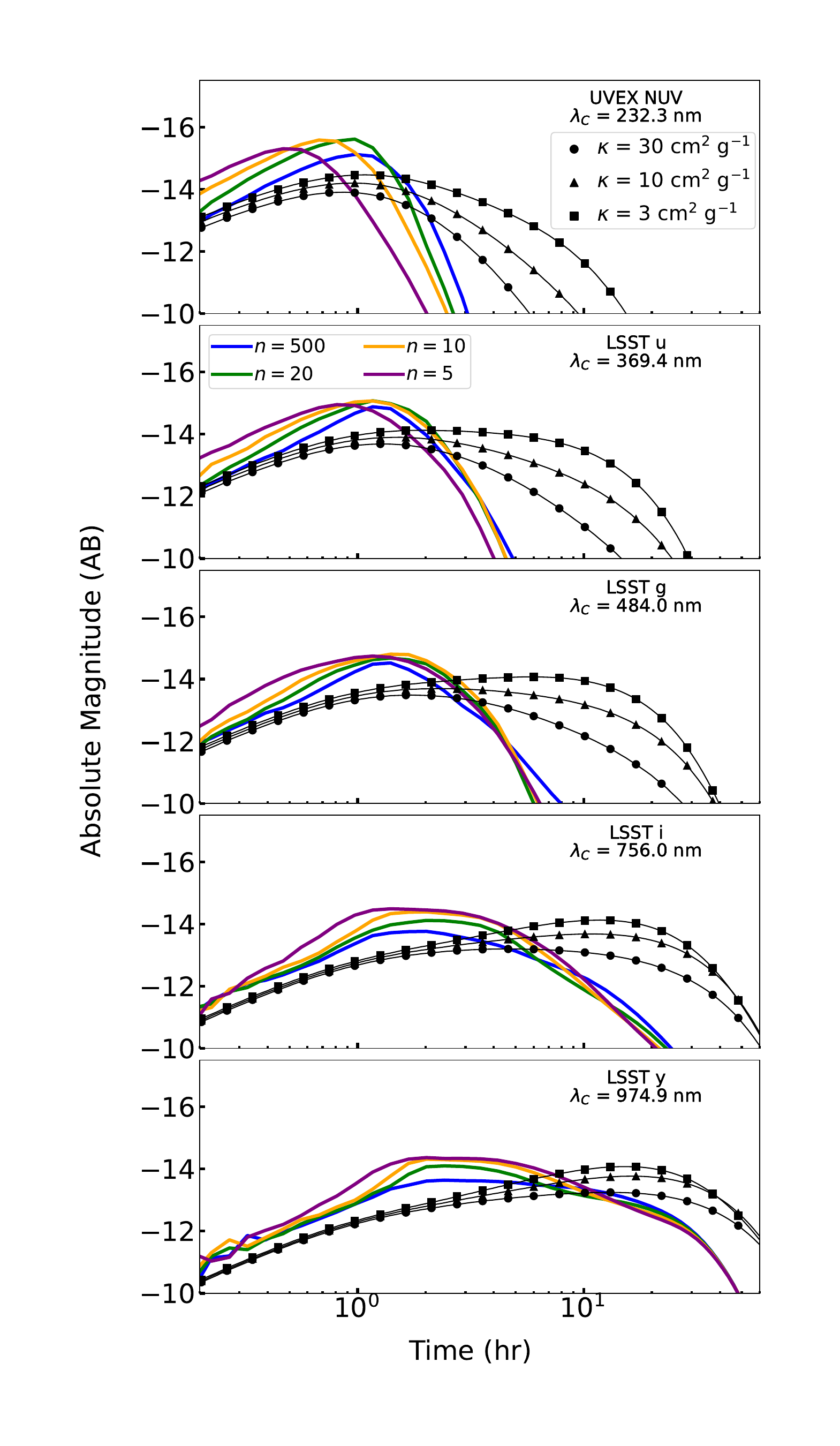}
    \caption{Comparison between the  analytical model of free neutron decay \citep{Metzger15} and the radiative transfer simulations from this work for M = $3\times10^{-3}$ M$_\odot$, of which M$_n$ = 10$^{-4}$ M$_\odot$, and variable mixing. The analytical model uses \vmax\,= $c$. The post-peak evolution in our models exhibits a much more rapid decay than the analytical model, and is most similar to gray opacity values of 10--30 cm$^2$ g$^{-1}$.}
    \label{fig:MetzgerComp}
\end{figure}

To compare to our model with $M_n = 10^{-4}$ M$_\odot$ that has an outer density profile power-law index of $-10$, we find that using the \cite{Metzger15} analytical form with $\bar{m} = 3\times10^{-3}$ M$_\odot$, $m_n$ = $3.25\times10^{-5}$ M$_\odot$, and $\beta = 7$ reproduces the same ejecta mass, neutron mass, and density power law. Figure \ref{fig:MetzgerComp} shows the light curves in UVEX NUV, $u, g, i,$ and $y$ bands of varying gray opacity analytical models against our radiative transfer models of the same properties and variable mixing. We find that there is no one analytical model that can fit all bands shown simultaneously, similarly to \cite{Banerjee24} for neutron-free models. The rise would require a gray opacity from $r$-process material that is less than the electron scattering opacity to match the brightness at optical wavelengths. The analytical model with $\kappa=3$ cm$^2$ g$^{-1}$ most closely matches the rise of our radiative transfer simulations at UVEX NUV and $u$ band, but the peak magnitude and peak timescale are not well-fit likely due to the rapid decline post-peak. The more rapid decline in our simulation is likely a result of the additional opacity from recombining $r$-process elements which is not captured in the analytical modeling. As a result of the different opacities, the peak magnitude in our radiative transfer simulations tends to be brighter by between 0.5--1 mag and peak earlier by at most a factor of $\sim2$ at UV wavelengths and $\sim5$ at optical wavelengths.

Beyond the optical light curves, the \cite{Metzger15} analytical model does not predict the emission of the X-rays that we see in our simulations likely as a consequence of averaging out the blackbody temperature over each zone in the model. Similarly, the model from \cite{Gottlieb20} predicts that there should be a gamma ray and/or X-ray signal from jet-ejecta interaction that produces a hot cocoon but not the free neutrons.

\begin{figure*}
    \centering
    \includegraphics[width=0.97\linewidth]{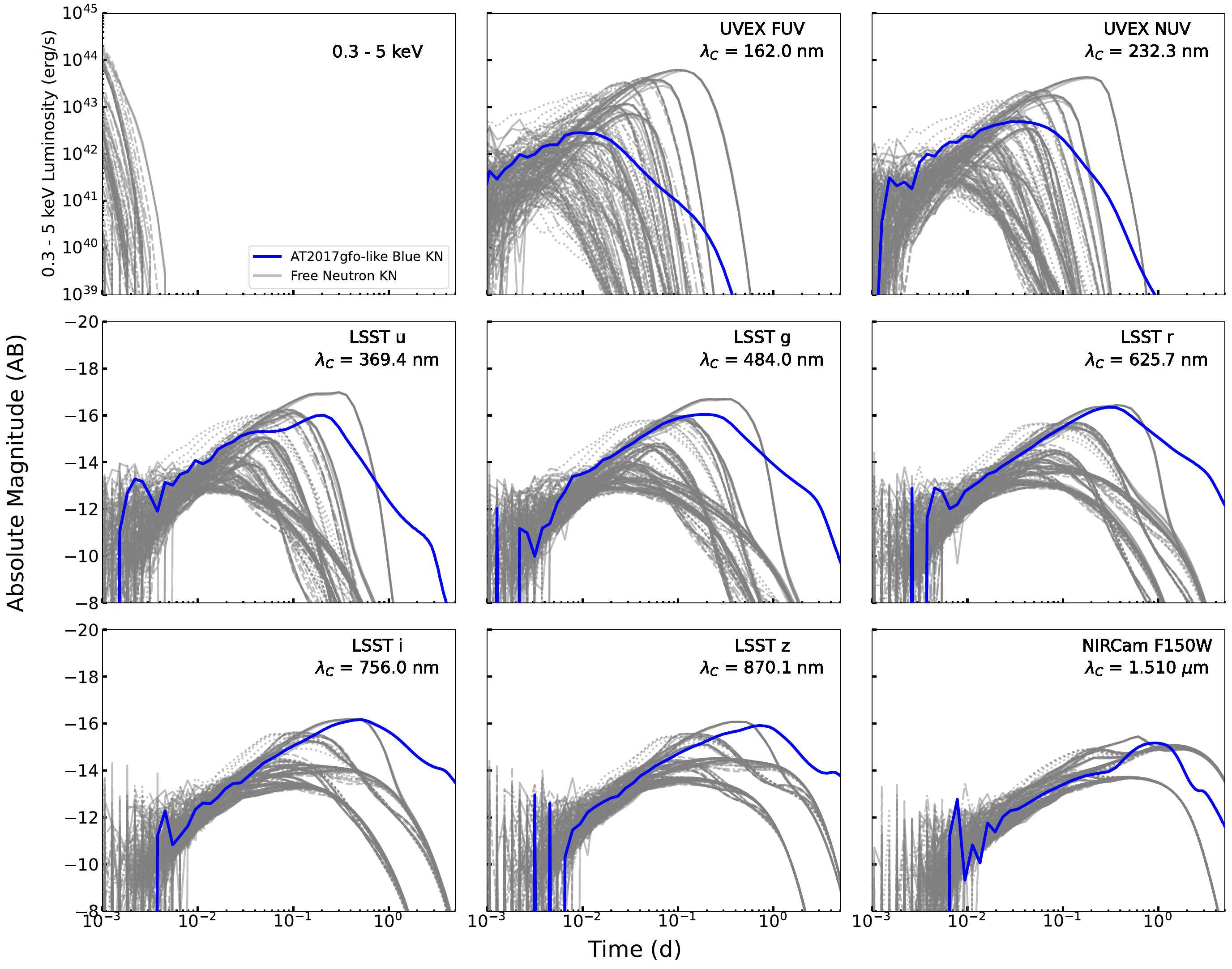}
    \caption{Representative light curve of an AT2017gfo-like blue kilonova component (blue line) with M$_{ej} = 2\times10^{-2}$ M$_\odot$, \Xlan\, = 0, and $v = 0.3c$ against all of the free-neutron models in our grid in gray. The blue KN component evolves much more slowly than models with free neutrons, and is much fainter in the X-rays and UV. While the optical peak magnitudes are similar to the most massive free neutron models, simultaneous observations in the UV can help to distinguish the blue component from the presence or lack of the associated UV peak. Monte Carlo noise is most prominent at early times and at redder wavelengths due to the short time bin width and weak red emission.}
    \label{fig:BlueComp}
\end{figure*}

\subsection{Comparison to Blue Kilonova Component}
\label{Subsec:Blue}
While free neutrons have been shown to produce ample early-time emission, disentangling the emission from other components of the KN ejecta may be difficult. To explore the uniqueness of free neutron emission, we compare our grid of models to a representative AT2017gfo-like blue KN model with M$ = 2\times10^{-2}$ M$_\odot$, \Xlan\, = 0, and $v_k = 0.3c$ (e.g., \citealt{Villar17, Rastinejad25}; though see complications of systematic errors in mass estimates in \citealt{Brethauer26}) from X-rays to NIR in Figure \ref{fig:BlueComp}. At optical and IR wavelengths, the models with the largest free neutron masses have comparable peak magnitudes to the blue component, but exhibit a much faster light curve evolution. All of the free neutron models fade much more rapidly than the AT2017gfo-like blue KN component, making the free neutron emission more distinguishable even if only caught at optical wavelengths. 

Simultaneous observations in the X-rays or UV with optical would further clarify whether the emission arises from free neutrons or the blue component. To produce the blue component optical peak magnitude and peak timescale from a free neutron model would require $M_n \gtrsim 10^{-3}$ M$_\odot$, which would produce a much brighter UV and X-ray luminosity than is observed in the blue component alone. UV observations are likely to be more constraining, as the UV emission is less sensitive to \vmax\, and mixing compared to the X-ray emission.

Additionally, from GRMHD simulations, radioactive heating from disk-wind ejecta could be responsible for the blue component (e.g., \citealt{Dessart09,Banerjee20,Curtis24}). This type of ejecta can be more massive and have typical velocities around $\sim0.1c$, which would help to further differentiate free neutron emission from the blue kilonova component as the longer photon diffusion timescale would result in a more fainter, redder, and more delayed peak. If both a free neutron component and a disk-wind component are present, the faster velocity required of the free neutron component may place the photosphere above the disk-wind ejecta at some viewing angles and suppress emission from the disk-wind, allowing the UV emission from the free neutrons to become observable over the blue component.

\vspace{-0.6cm}
\subsection{Opacities} \label{Subsec:Opac}

\begin{figure*}
    \centering
    \includegraphics[width=0.97\linewidth,trim={2cm 2.5cm 2.6cm 2.5cm},clip]{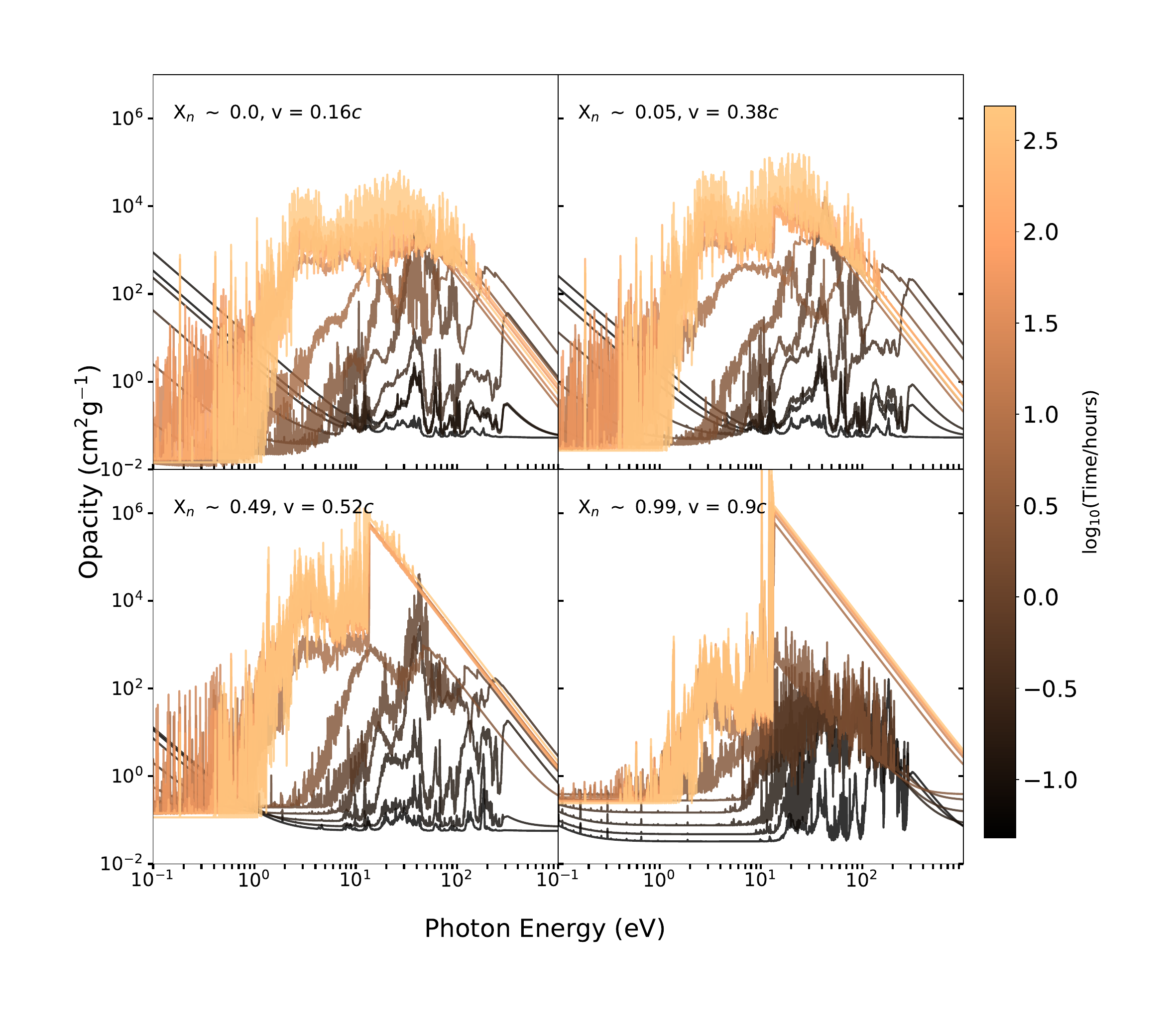}
    \caption{Time evolution of the wavelength-dependent opacity in representative neutron-free to neutron-rich layers of the ejecta for the $M_n = 10^{-4}$ M$_\odot$, $v_{\rm max} = 0.9c$, $n = 10$ model. Much of the opacity at $t \lesssim 1$ hr is dominated by a combination of bound-bound and bound-free from the $r$-process species, even in zones where there is very little $r$-process mass.}
    \label{fig:Opac}
\end{figure*}

\begin{figure}
    \centering
    \includegraphics[width=0.97\linewidth,trim={2cm 2.5cm 2.6cm 2.3cm},clip]{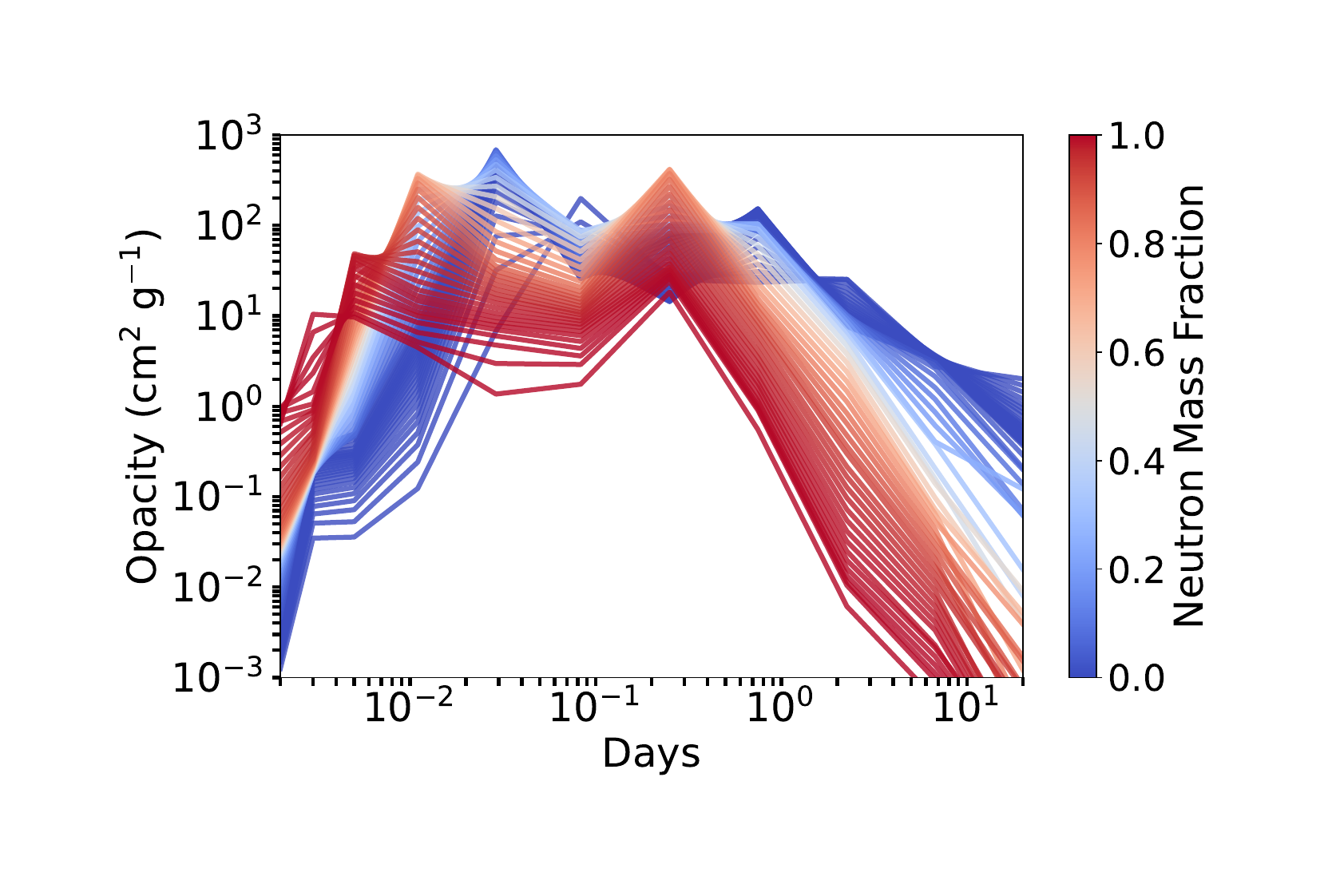}
    \caption{Time evolution of the Planck mean opacity for each zone in the M$_n$ = 10$^{-4}$ M$_\odot$, $n = 500$ model as a function of the initial free neutron mass fraction. At $t \lesssim 1$ day, the opacity is well-characterized across most zones by $\kappa \sim 10-100$ cm$^2$ g$^{-1}$ with lower $X_n$ zones typically having smaller Planck mean opacities. However, at $t \gtrsim 1$ day, the opacity in high $X_n$ zones rapidly declines below }
    \label{fig:Planck}
\end{figure}

While full radiative transfer simulations are more realistic, the high computational cost can be prohibitive in some scenarios. In these cases, we detail the opacities calculated in the neutron-rich layers of the ejecta and the bulk dynamical ejecta that can be utilized in smaller scale radiative transfer simulations using the output of the M$_n$ = 10$^{-4}$ M$_\odot$, $n = 10$, \vmax\, = 0.9$c$ model.

Initially, when the ejecta is highly ionized the opacity at blue wavelengths is dominated by electron scattering while the opacity at red wavelengths is dominated by free-free (Figure \ref{fig:Opac}). Electron scattering can dominate when the ions have been stripped down to d-, p-, and s-shell orbitals that contribute roughly 1, 0.1, and 0.01 cm$^2$ g$^{-1}$ to the opacity, respectively \citep{Banerjee24}. If higher Z elements than used in this work are present and are stripped to have valence f-shell electrons, those elements can dominate the line expansion opacity. As the ejecta evolve, within $t \lesssim 1$ hr bound-bound and bound-free opacities of $r$-process elements dominate at blue wavelengths even when the ejecta is almost entirely comprised of hydrogen. As the ejecta cool, eventually $\sim50\%$ of the freshly synthesized hydrogen recombines to neutral and generates significant bound-free opacity blueward of 13.6 eV, rapidly overtaking bound-bound opacity. For smaller M$_n$, the hydrogen becomes neutral on a shorter timescale and dominates the opacity sooner than $\sim 1$ day in the case of M$_n$ = 10$^{-4}$ M$_\odot$.

Figure \ref{fig:Planck} shows the evolution of the Planck mean opacity as a function of time across ejecta zones as a function of initial free neutron mass fraction. Initially, the zones with $X_n \gtrsim 0.5$ show a higher and increasing Planck mean opacity, reaching $\sim \rm{few} \times 100$ cm$^2$ g$^{-1}$ with typical values of $\sim \rm{few} \times 10$ cm$^2$ g$^{-1}$ before the ejecta cool and the opacity decreases at $t \gtrsim 1$ day. Similarly, at $t \gtrsim 1$ day, zones with higher $X_n$ lack of $r$-process elements and so have lower Planck mean opacity than those with significant $r$-process elements.

\vspace{-0.6cm}
\subsection{Impact of Dynamical Ejecta Mass}
\label{App:Red}

\begin{figure}
    \centering
    \includegraphics[width=0.97\linewidth,trim={0.1cm 0.3cm 0cm 0cm},clip]{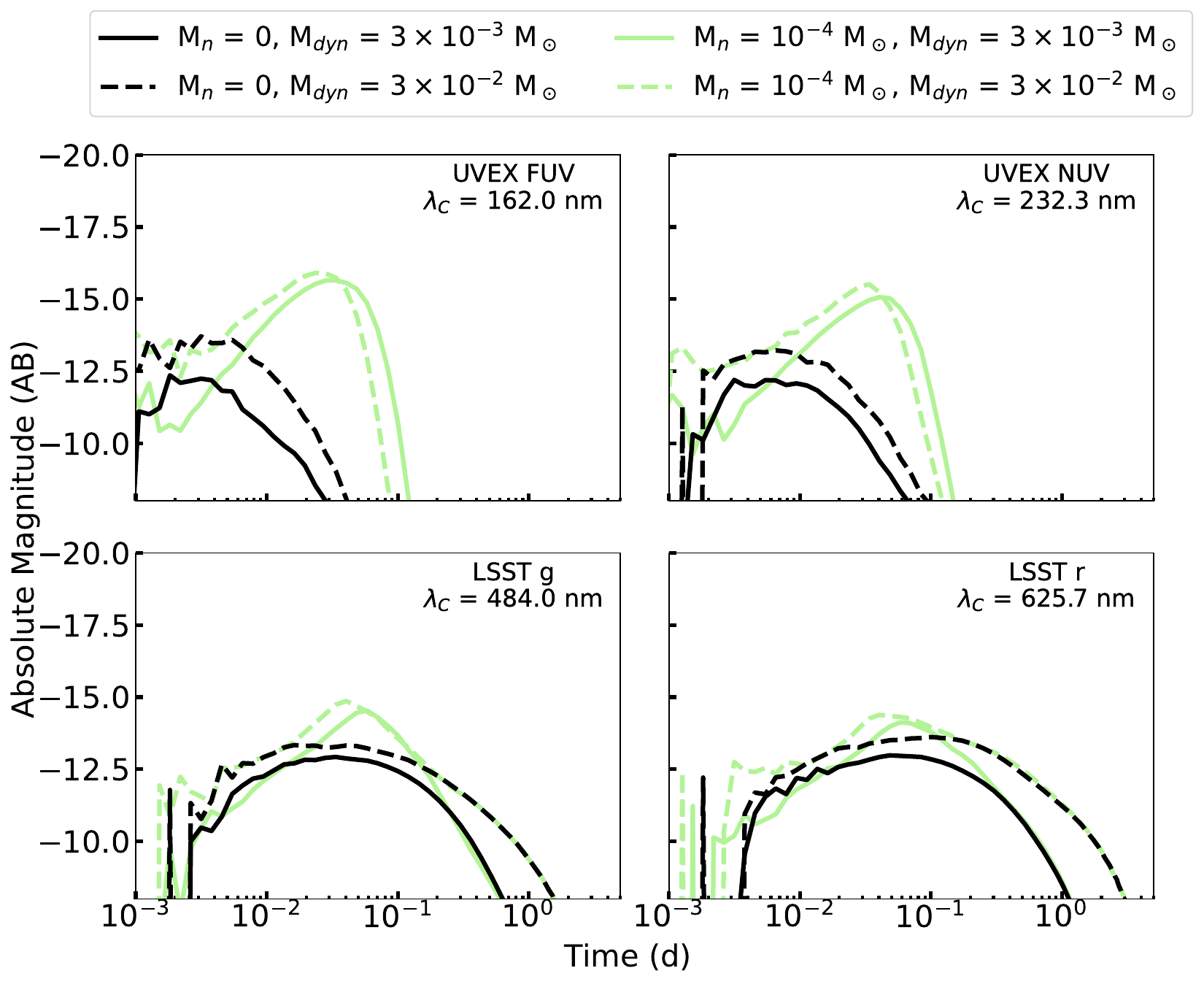}
    \caption{Comparison of models with $M_n$ = $10^{-4}$ M$_\odot$ and without free neutrons for underlying dynamical ejecta masses of $3\times10^{-3}$ and $3\times10^{-2}$ M$_\odot$ (solid and dashed lines, respectively). }
    \label{fig:RedComp}
\end{figure}

We explore the impact of dynamical ejecta mass on the emission from free neutrons through a more massive AT2017gfo-like red component of M = $3\times10^{-2}$ M$_\odot$, $v_k = 0.3c$, and \Xlan= 0.1 that contains $M_n = 10^{-4}$ M$_\odot$ with $n = 500$ and \vmax = $0.7c$. The higher mass dynamical ejecta leads to a brighter transient at later times when the bulk of the $r$-process ejecta dominate the light curve, but have nearly identical free neutron-dominated light curves (Figure \ref{fig:RedComp}). At UV wavelengths, the peak timescale is notably earlier at $\sim 0.03$ days in the more massive dynamical ejecta compared to $\sim0.04$ days in the lighter dynamical ejecta. The difference in peak timescale is similar at optical wavelengths, also showing the more massive dynamical ejecta is brighter leading up to the peak. The brighter rise may be a result of the photosphere residing farther out in the more massive dynamical ejecta where there is a higher neutron fraction and therefore a more luminous and hotter spectral evolution. The earlier peak timescale in the UV could be due to the outermost ejecta becoming neutral on a faster timescale from the higher densities and higher recombination rates. Once the outermost layers begin to turn neutral, the $r$-process material rapidly blankets the UV in a forest of lines that causes the light curve to begin a rapid decline.

\vspace{-0.6cm}
\subsection{Modeling Uncertainties}

\label{subsec:limits}

Here, we briefly discuss the modeling uncertainties that are intrinsic to our models and their potential impacts on the observables. See \cite{Brethauer26} for modeling uncertainties specific to the implementation of QNLTE.

\begin{itemize}

    \item \textbf{Density Profiles} - At $t \lesssim 1$ day, the location of the photosphere, and thus the resulting emission, is strongly influenced by the precise density profile chosen. High energy emission such as X-rays and UV are particularly sensitive to the mixing, as shown by Figure \ref{Fig:SpecSeqDensVel}. While we chose to model the density profile as a broken power-law with $\delta = -1$ and $q = -10$, models with shallower density profiles may produce more X-ray and UV emission at ($t \lesssim 1$ hr) as a result of a photosphere radius at the edge of the ejecta for a longer duration and less absorbing material for the photons to travel through. Once the photosphere begins to recede into the ejecta, shallower density profiles may produce a fainter UV transient from higher absorption. 

    \item \textbf{Recombination Rates} - Dielectronic recombination is expected to dominate the recombination rate for some $r$-process species (e.g., \citealt{Hotokezaka21,Banerjee25,Singh25}). However, experimental and theoretical estimates of dielectronic recombination rates for $r$-process species are sparse, especially for highly ionized species that are important at $t \lesssim 1$ day. If recombination rates are higher than the values used in this work (see \citealt{Brethauer26} for further details), this could lead to a suppression of blue emission as seen in \cite{Brethauer26} with higher recombination rates.

    \item \textbf{Multi-dimensional Ejecta} - The models we present here are all spherically symmetric, though there may be viewing angle effects. Free neutrons are most common in high velocity and high entropy ejecta \citep{Schnabel26}, which can be realized by the shock-heated ejecta and therefore have a quasi-spherical geometry \citep{Bauswein13,Radice18}. This would lead to a higher concentration of free neutrons along a preferred axis, and could produce a brighter X-ray or UV flash along some viewing angles while suppressing along others from the change in visible surface area composed of free neutrons.

\end{itemize}
\vspace{-0.8cm}
\section{Conclusions} \label{Sec:Conc}

For the first time, we presented quasi-NLTE radiative transfer simulations of free neutrons embedded within the high-velocity tail ($\gtrsim 0.4c$) of kilonova ejecta that can power X-ray and UV transients lasting up to $\sim$minutes and $\sim$ hours, respectively, detectable by observational facilities like UVEX, ULTRASAT, and Einstein Probe. The duration and peak magnitude of the signals are strongly dependent on the mixing of the free neutrons with the underlying dynamical ejecta, the total mass of free neutrons, and the maximum velocity that the ejecta extend to. 

First, the X-rays from free neutrons can be distinguishable from other emission mechanisms by the soft X-ray energies, $\sim$ minute duration, and typical luminosities of 10$^{41} - 10^{43}$ erg s$^{-1}$. Following the X-ray flash, the FUV and NUV are uniquely capable of revealing the presence of even small amounts of free neutrons, down to $\sim 10^{-7}$ M$_\odot$ with a distinctive excess in the FUV during the first $\sim 30$ minutes post explosion. The UV spectra at early times ($t \lesssim 2$ hrs) can show deviations from a blackbody that are dependent on ejecta properties. However, the presence of free neutrons can counterintuitively result in more rapidly declining and fainter UV/optical emission as the freshly synthesized protons absorb some of the ionizing $\beta$-decay energy from the $r$-process material, causing the overall ionization state of the outermost ejecta to decrease and become more opaque.

Compared to previous analytical models, the full radiative transfer simulations tend to be brighter, fade more rapidly, and may be able to produce an X-ray flash. Despite the mismatch in analytical model and radiative transfer simulations, the peak magnitudes are broadly within 1 mag of our radiative transfer simulations, though the analytical models tend to peak at later times by a factor of $\sim2-5$. The shape of the light curve is not well matched, but an $r$-process gray opacity initially of $\sim$3 cm$^2$ g$^{-1}$ which then transitions to $\sim$10--30 cm$^2$ g$^{-1}$ post-peak, depending on the photometric band, can approximately capture the decline rates. 

Further studies of the high-velocity tail of ejecta will be critical to improving radiative transfer simulations. Characterizing the density profile, mass, and composition of the high-velocity tail  would allow for the production of higher-resolution grids of models, and better enable characterization of the nucleosynthetic environment of neutron star mergers.

\vspace{-0.6cm}
\section*{Acknowledgements}

Research at UC Berkeley is conducted on the territory of
Huichin, the ancestral and unceded land of the Chochenyo speaking Ohlone people, the successors of the sovereign
Verona Band of Alameda County.

High-end computing (HEC) resources supporting this work were provided by the NASA Advanced Supercomputing (NAS) Division at Ames Research Center.

DB is partially supported by a NASA Future Investigators in NASA Earth and Space Science and Technology (FINESST) award No. 80NSSC23K1440 and the Hearts to Humanity Eternal Research Grant through UC Berkeley. R.M. acknowledges support by the National Science Foundation under award No.  AST-2224255. %The TReX team at UC Berkeley is partially funded by the Heising-Simons Foundation under grant 2021-3248 (PI: Margutti).
DK is supported in part by the U.S. Department of Energy, Office of Science, Division of Nuclear Physics, under award numbers DE-SC0004658 and DE-SC0024388, and by the Simons Foundation (award number 622817DK). 

SB acknowledges the support from the European Union’s Horizon Europe research and innovation program under the Marie Skłodowska-Curie grant agreement No. 101274945.

%%%%%%%%%%%%%%%%%%%%%%%%%%%%%%%%%%%%%%%%%%%%%%%%%%
\vspace{-0.6cm}
\section*{Data Availability}

All spectra will be made publicly available through Zenodo.

\textit{Software:} numpy \citep{Numpy}, \texttt{sedona} \citep{Kasen06,Roth15}, 
%astropy \citep{astropy:2013,astropy:2018,astropy:2022},
matplotlib \citep{Matplotlib}, h5py

%%%%%%%%%%%%%%%%%%%% REFERENCES %%%%%%%%%%%%%%%%%%

% The best way to enter references is to use BibTeX:
\vspace{-0.6cm}
\bibliographystyle{mnras}
\bibliography{KNe} % if your bibtex file is called example.bib

% Alternatively you could enter them by hand, like this:
% This method is tedious and prone to error if you have lots of references
%\begin{thebibliography}{99}
%\bibitem[\protect\citeauthoryear{Author}{2012}]{Author2012}
%Author A.~N., 2013, Journal of Improbable Astronomy, 1, 1
%\bibitem[\protect\citeauthoryear{Others}{2013}]{Others2013}
%Others S., 2012, Journal of Interesting Stuff, 17, 198
%\end{thebibliography}

%%%%%%%%%%%%%%%%%%%%%%%%%%%%%%%%%%%%%%%%%%%%%%%%%%

%%%%%%%%%%%%%%%%% APPENDICES %%%%%%%%%%%%%%%%%%%%%

%\appendix

%\section{}

%If you want to present additional material which would interrupt the flow of the main paper,
%it can be placed in an Appendix which appears after the list of references.

%%%%%%%%%%%%%%%%%%%%%%%%%%%%%%%%%%%%%%%%%%%%%%%%%%

% Don't change these lines
\bsp	% typesetting comment
\label{lastpage}
\end{document}